\documentclass[review]{elsarticle}
\usepackage[margin=2.54cm]{geometry}
\usepackage{tikz}
\usetikzlibrary{decorations.pathreplacing}
\usepackage{framed}
\usepackage{mdframed}
\usepackage[most]{tcolorbox}
\usepackage{colortbl}
\usepackage{listliketab}
\usepackage{listings}
\makeatletter
\def\ps@pprintTitle{%
	\let\@oddhead\@empty
	\let\@evenhead\@empty
	\def\@oddfoot{}%
	\let\@evenfoot\@oddfoot}
\makeatother

\usepackage{lineno,hyperref}
\usepackage[version=3]{mhchem}
\usepackage{amssymb, amsmath, amsthm, graphicx,booktabs, float, multirow}
\usepackage{multirow}
\usepackage{subcaption}
\usepackage{caption}
\usepackage[makeroom]{cancel}
\usepackage{rotating}
\usepackage{xparse}
\NewDocumentCommand{\ceil}{s O{} m}{%
	\IfBooleanTF{#1} 
	{\left\lceil#3\right\rceil} 
	{#2\lceil#3#2\rceil} 
}

\usepackage[labelfont=bf]{caption}

\modulolinenumbers[5]

\usepackage{array}
\newcolumntype{P}[1]{>{\centering\arraybackslash}p{#1}}
\newcolumntype{M}[1]{>{\centering\arraybackslash}m{#1}}

\newtheorem{theorem}{Theorem}[section]

\newtheorem{algorithm}[theorem]{Algorithm}

\usepackage{xcolor}

\definecolor{codegreen}{rgb}{0,0.6,0}
\definecolor{codegray}{rgb}{0.5,0.5,0.5}
\definecolor{codepurple}{rgb}{0.58,0,0.82}
\definecolor{backcolour}{rgb}{0.95,0.95,0.92}

\lstdefinestyle{mystyle}{
	backgroundcolor=\color{backcolour},   
	commentstyle=\color{codegreen},
	keywordstyle=\color{magenta},
	numberstyle=\tiny\color{codegray},
	stringstyle=\color{codepurple},
	basicstyle=\ttfamily\footnotesize,
	breakatwhitespace=false,         
	breaklines=true,                 
	captionpos=b,                    
	keepspaces=true,                 
	numbers=left,                    
	numbersep=5pt,                  
	showspaces=false,                
	showstringspaces=false,
	showtabs=false,                  
	tabsize=2
}

\usepackage{tcolorbox}
\tcbuselibrary{theorems}

\newtcbtheorem[number within=section]{mytheo}{Algorithm}%
{colback=green!5,colframe=green!35!black,fonttitle=\bfseries}{th}

\begin{document}
	\begin{frontmatter}
\title{Entropic Measures of Complexity in a New Medical Coding System}


\author[mymainaddress]{Jerome Niyirora, PhD\corref{mycorrespondingauthor}}
\ead{jerome.niyirora@sunypoly.edu}
\cortext[mycorrespondingauthor]{Corresponding author. Tel.:+0113157927430}
\address[mymainaddress]{SUNY Polytechnic Institute, College of Health Sciences, Utica, New York, USA}
\begin{abstract}
	\section*{Background}
		Transitioning from an old medical coding system to a new one can be challenging, especially when the two coding systems are significantly different. The US experienced such a transition in 2015.
	\section*{Objective}
		This research aims to introduce entropic measures to help users prepare for the migration to a new medical coding system by identifying and focusing preparation initiatives on clinical concepts with more likelihood of adoption challenges.		
	\section*{Methods}	
		Two entropic measures of coding complexity are introduced. The first measure is a function of the variation in the alphabets of new codes. The second measure is based on the possible number of valid representations of an old code. 
	\section*{Results}
		 A demonstration of how to implement the proposed techniques is carried out using the 2015 mappings between ICD-9-CM and ICD-10-CM/PCS. The significance of the resulting entropic measures is discussed in the context of clinical concepts that were likely to pose challenges regarding documentation, coding errors, and longitudinal data comparisons.
	\section*{Conclusion}		 
		The proposed entropic techniques are suitable to assess the complexity between any two medical coding systems where mappings or crosswalks exist. The more the entropy, the more likelihood of adoption challenges. Users can utilize the suggested techniques as a guide to prioritize training efforts to improve documentation and increase the chances of accurate coding, code validity, and longitudinal data comparisons. 
\end{abstract}

\begin{keyword}
	Medical Coding \sep Mapping \sep Complexity \sep Entropy 
\end{keyword}

\end{frontmatter}
\section*{Key messages}
\begin{enumerate}
	\item Research suggests that Shannon's entropy can be applied to estimate the complexity of transitioning from an old medical coding system to a new one. 
	\item The more the entropy, the more likelihood of transition challenges.
	\item Users can utilize the proposed techniques as a guide to prioritize training efforts to improve clinical documentation and increase the chances of accurate coding, code validity, and longitudinal data comparisons. 
\end{enumerate}

\section*{Declarations}
\subsection*{Funding}
No funding was received to conduct this study.
\subsection*{Conflicts of interest/Competing interests}
None
\subsection*{Availability of data and material}
Yes
\subsection*{Code availability}
Yes
\subsection*{Authors' contributions}
There is only one author for this manuscript

\newpage
\section{INTRODUCTION} \label{sec_intro}
Medical diagnoses and procedures are reported using standardized codes that are updated periodically to keep up with the latest clinical knowledge and practices. Transitioning from an old medical coding system to a new one can be challenging, especially when the two systems are significantly different. One such transition took place in the United States (US) in 2015 when the country switched from the $9^{th}$ revision of the International Classification of Diseases (ICD) Clinical Modification (ICD-9-CM) to the $10^{th}$ revision (ICD-10-CM). This newer revision was accompanied by a very different procedure coding system (PCS) (ICD-10-PCS), as compared to the ICD-9-CM procedure coding system (Volume 3, abbreviated here as Vol. 3). For example, each ICD-10-PCS procedure is made of 7 multi-axial characters where each axis encompasses up to 34 alphanumeric values \citep{cms_guidelines}. This arrangement is a significant departure from the procedure code structure in ICD-9-CM Vol. 3, where all codes are numeric and can only be between 2 and 4 characters long. In 2015, ICD-10-PCS had about 72,000 procedure codes as compared to only about 4,000 codes in ICD-9-CM Vol. 3. The diagnosis codes between these two revisions of ICD are also quite different. For example, all diagnosis codes in ICD-10-CM are alphanumeric and can be 3 to 7 characters long, whereas ICD-9-CM diagnosis codes are mostly numeric and can only be between 3 and 5 characters long. In 2015, there were about 14,500 diagnosis codes in ICD-9-CM as compared to about 69,800 codes in ICD-10-CM  \citep{nchs}. Given these differences, some analysts had predicted a costly and challenging transition from ICD-9-CM to ICD-10-CM/PCS \citep{caskey2014transition}. 
Indeed, some of the feared problems did materialize after the changeover, such as the loss in productivity \citep{kusnoor2019narrative, alakrawi2017new}, the lack of readiness of computer systems, the inability to find some ICD-9-CM concepts in the ICD-10-CM system, and difficulties mapping ICD-10-CM to other coding systems such as SNOMED-CT \citep{monestime2019analyzing}. Some ICD-10-CM clinical classes were also found to have more coding deficiencies than others, such as the class of external causes of morbidity (V00-Y99) \citep{butler2016analyzing}. In one post-ICD-10 implementation audit, it was found that one of the most significant challenges for coders was selecting the correct character in the 3rd position (Root Operation), the 4th position  (Body Part), and the 5th position (Approach) of an ICD-10-PCS code \citep{Stitcher}. While little evidence exists to suggest that reimbursement was significantly impacted by the transition, in some practices, a statistical increase in the coding-related denials was noted \citep{hellman2018impact}. A few of the post-transition qualitative studies concluded that training and education were critical in overcoming many of the previously anticipated challenges \citep{monestime2019analyzing, krive2015complexity}. Besides the US, other countries have also faced challenges while transitioning to new medical coding systems. The issues ranged from coding errors to discrepancy problems when the same condition was coded in both coding systems. For example, in one analysis \citep{januel2011improved}, it was found that the Swiss transition from ICD-9 to ICD-10 resulted in the initial increase of the number of coding errors for co-morbidities, but, over time, the accuracy improved as the learning curve waned. In one Canadian study \citep{quan2008assessing}, the authors were interested in assessing the validity of ICD-10 codes after switching from ICD-9. While the authors did not find much difference in the validity of the codes from these two systems, the discrepancy was apparent for some conditions (e.g., HIV/AIDS, hypothyroidism, and dementia). The authors also observed that the quality of data had not yet improved in ICD-10 as originally expected. 

Now that many countries are preparing to migrate from ICD-10 to ICD-11 \citep{who_icd11}, one can expect similar transition challenges to occur, as these two coding systems have different code structures \citep{cdc_icd11}, and the equivalence is at times lacking \cite{fung2020new}. This research aims to introduce entropic measures to help users prepare for the migration to a new medical coding system by identifying and focusing preparation initiatives on clinical concepts with more likelihood of documentation deficiencies, coding errors, and longitudinal data comparison issues.

\section{RELATED WORK}
Not many studies have considered how to quantify the complexity of codes between two medical coding systems. In some studies, the equivalence in the number and structure of the codes between two coding systems is considered, but without accompanying measures of the dissimilarity in the codes \cite{fung2020new}. In a few studies, an attempt is made to address the complexity between two medical coding systems. For example, in Boyd et al. \cite{boyd2013discriminatory, boyd2018icd}, the authors proposed using the science of networks to evaluate the difficulties of transitioning from ICD-9-CM to ICD-10-CM in the US. The authors used General Equivalence Mappings (GEMs) to create graphs where diagnoses were nodes, and the relationships in the GEMs were edges. From their analysis, the authors derived directional motifs and identified convoluted mappings, where multiple medical codes from both coding systems shared complex, entangled, and non-reciprocal mappings. The authors concluded that clinical classes with convoluted mappings were more likely to be challenging to code and costly to implement after the changeover to the new medical coding system. Besides, these authors also anticipated that clinical classes with a high ratio of ICD-10-CM to ICD-9-CM codes were more likely to affect a smooth transition.
Another study that considered the complexity of transitioning between two coding systems relates to the work of Chen et al. \cite{chen2018leveraging}, where the authors leveraged  Shannon's entropy to develop a mapping framework between ICD-10 and ICD-11 coding systems. The authors proposed three entropy-based metrics of standardizing rate (SR), uncertainty rate (UR), and information gain (IG) to validate information changes between ICD-10 and ICD-11. The authors obtained the UR measure by $\sum_{i=1}^{M} p_i\log 1/p_i$, where $M$ was the number of ICD-11 candidate codes for a single ICD-10 code, and $p_i$ was the probability of each ICD-11 code. In a special case of a uniform distribution, the authors suggested utilizing the average probability of $1/M$ to measure $UR$, which implied that $UR = \log M$. Among other conclusions, the authors recommended verifying ICD-10 codes with high UR measures as these codes were more likely to hinder a smooth transition to ICD-11.

\section{CONTRIBUTIONS}
This research complements previous studies highlighted in the Related Work section. For example, as in Chen et al. \cite{chen2018leveraging}, this research proposes to apply Shannon's entropy to study the complexity of the transition between two medical coding systems. Unlike in this previous study, the entropic measures in this research account for the variation in the alphabets of candidate codes. Besides, Shannon's entropy is also used to create a measure of coding complexity that considers not only the number of candidate codes (as in the UR measure \cite{chen2018leveraging}) but also the number of combinations of these codes. As shown later, failure to account for the latter information may underestimate or overestimate the related coding complexity. It should also be mentioned that the proposed methods have an advantage over convoluted measures suggested in Boyd et al. \cite{boyd2013discriminatory, boyd2018icd}. Unlike in the convoluted approach, where a code is classified as either being involved in a convoluted relationship or not,  the proposed methods provide non-dichotomous complexity measures of each code.

\section{MATERIALS AND METHODS} \label{sec_methods}
\subsection{Methods}
\subsubsection{A motivating problem}
It is imagined that a manager of a given medical care facility is preparing to transition from an old medical coding system $X$ to a new medical coding system $Y$. The forward ($X\rightarrow Y$) and backward ($X\leftarrow Y$) mappings between $X$ and $Y$ are provided. The manager is unsure about employing these mappings to identify clinical concepts that are more likely to be challenging to translate into the new medical coding system. Some of the benefits of knowing this information include being able to formulate targeted training efforts for coding and clinical documentation to foster the validity of the data in the new coding system. Besides, understanding complex translations may help take the necessary steps to ensure longitudinal data comparisons. This research aims to suggest the techniques that the manager could use to solve this dilemma.

\subsubsection{Model and assumptions}
Given forward mappings ($X\rightarrow Y$), the old medical coding system $X$ is termed the source system, while the new coding system $Y$ is termed the target system. In the backward mappings ($X\leftarrow Y$), the \textit{source} and \textit{target} terminologies are reversed. For model development, only forward mappings are considered here since the backward mappings would obey the same logic. From the prescribed forward mappings ($X\rightarrow Y$), it is assumed that code $x\in X$  corresponds to $m$ number of candidate codes $y \in Y$. This relationship, referred to here as a \textit{map}, is symbolized as $x\rightarrow \left\{y_1, y_2,\dots,y_m\right\}$ or as in the following matrix form:
\begin{eqnarray} \label{matrix_A}
x\rightarrow
\begin{bmatrix}
a_{11} & a_{12} & \dots& a_{1n} \\
a_{21} & a_{22} & \dots& a_{2n} \\
\dots &\dots & \dots& \dots \\
a_{m1} & a_{m2} & \dots& a_{mn}
\end{bmatrix}
=
\begin{bmatrix}
	y_{1} \\
	y_{2} \\
	\dots  \\
	y_{m} 
\end{bmatrix}
\end{eqnarray}
where each code in the map $y_i$, for $i:1,\dots,m$, has $n$ fixed number of characters (also called alphabets) $a_{ij}$, for $j:1,\dots,n$. If necessary, padding may be added to a particular code to ensure that $n$ is fixed for all codes, an approach, as shown later, simplifies calculations. Each column represents an axis or simply a position of an alphabet in a code. The columns of a map are assumed independent. Each row of a map represents a valid code $y\in Y$. A set of more than one code in a map may be necessary to represent code $x \in X$. If $m = 0$, code $x$ has no match in $Y$, which implies data loss in the new coding system. If $m = 1$, code $x\in X$ has a one-to-one relationship with code $y \in Y$. In this case, the coding complexity is expected to be zero since little surprise exists about what the new code should be. If $m > 1$, the coding complexity will be greater than zero as there is more than one candidate code in $Y$, thus more complexity and chances of coding or translation errors. In this research, a coding error is defined as the selection of a code where at least one alphabet is wrong or the selection of a set of codes where at least one of the codes is incorrect or missing. The expected coding complexity of a given clinical concept in $X$ is characterized in terms of the uncertainty in the rows and columns of a map, which is measured here in bits units of Shannon's entropy \citep{luenberger}.

Two major sources of coding complexity are assumed here, namely source $A$, which captures the variation in the alphabets of a map, and source $B$, which relates to the combinations of the rows of a map. The entropy for source $A$, or $H(A)$, is calculated as:
\begin{eqnarray}\label{eq_H(A)}
H(A) =  -\sum_{j = 1}^{n}\sum_{i = 1}^{k_j} p_{ij}\log_2 p_{ij} \equiv \sum_{j=1}^{n} H(\boldsymbol{a_j})
\end{eqnarray} 
where $k_j\leq m$ is the number of unique alphabets in column  $\boldsymbol{a_{j}}$ of matrix (\ref{matrix_A}) and $p_{ij}$ is the probability of alphabet $i$ in position $j$. The more the $H(A)$ measure, the more requisite detailed documentation to express all the alphabets of a map. Likewise, the more the number of code alphabets that must be chosen separately, the more complex and time-consuming the coding.

Regarding source $B$, the corresponding entropy $H(B)$ is obtained by:
\begin{eqnarray}\label{eq_H(B)}
H(B) =\log_2 (v)
\end{eqnarray}
where $v = m_0 + \sum_{i=1}^{s}\prod_{j = 1}^{m - m_0}m_{ij}$. Here, $s$ is the total number of possible scenarios and $m_0$ represents the number of stand-alone codes and, for a given scenario $i$, $m_{i1},\dots, m_{i(m-m_0)}$ denote the number of candidate codes in Y that must be combined to represent code $x\in X$. As before, $m$ is the total number of candidate codes in a map. If a map only includes stand-alone codes, where no combinations of codes are required, Equation \ref{eq_H(B)} becomes comparable to the UR measure introduced in Chen et al. \cite{chen2018leveraging}. The more the $H(B)$ measure, the more complex the coding due to the need for more coding memory and time, since more than one candidate code in the target system is going to be required to represent a single code from the source system. See \ref{appendix_entropy} for more details on the derivation of Equations \ref{eq_H(A)} and \ref{eq_H(B)}.

\subsubsection{Implementation}
It is recommended that both $H(A)$ and $H(B)$ entropic measures be normalized into $Z(\alpha)$ and $Z(\beta)$, as exemplified in \ref{appendix_entropy}, to allow for the comparison and ranking of complexity from different sources. If $H(A)$ and $H(B)$ measures (or their normalized counterparts) are to be utilized to prepare for the transition (e.g., documentation improvement), they should be weighed using relevant  empirical distribution (e.g., historical frequencies of codes in a given medical facility or general practice area). Accordingly, if, say, a particular facility never performs heart transplants, it shouldn't have to spend too much training efforts on the documentation of this clinical concept. Algorithm \ref{algorithm_shannon} shows the steps that one can take to implement the suggested entropic methods.
\begin{algorithm}\label{algorithm_shannon}
\begin{mytheo}{Computing entropic measures}{theoexample}
	\begin{description}
		\item [Step 1: ] Calculate $H(A)$, the entropy of the columns of a map, to estimate the coding complexity due to the variation in the alphabets of the columns of a map. 
		\item [Step 2: ] Calculate, $H(B)$, the entropy of the rows of a map to estimate the coding complexity due to the uncertainty in the number of valid code representations in the map. 
		\item [Step 3: ] Normalize $H(A)$  and $H(B)$ by centering these measures and then dividing them by their standard deviations. The normalized measures are symbolized here as $Z(\alpha)$ for $H(A)$ and $Z(\beta)$ for $H(B)$. 
		\item [Step 4: ] If empirical data, based on historical visits or future forecasts, were available, one would adjust $Z(\alpha)$ and $Z(\beta)$  measures by multiplying them with the probability of a corresponding clinical concept.
		\item [Step 5: ] Use the adjusted or unadjusted entropic measures to prioritize transition initiatives between two medical coding systems.	
	\end{description}
\end{mytheo}
\end{algorithm}

\subsection{Materials}
Algorithm \ref{algorithm_shannon} can be applied to evaluate entropic measures between any two medical coding systems, provided mappings or crosswalks exist. For demonstration, the 2015 US transition from ICD-9-CM to ICD-10-CM/PCS medical coding systems is considered. For a brief background, when the US was preparing to migrate from ICD-9-CM to ICD-10-CM/PCS, forward and backward general equivalence mappings (GEMs) were made available to users \citep{nchs, cms_gem}. A user could determine the number of candidate codes in the target system from these mappings, given a code in the source system. These files also allowed users to apply the given supplemental five digits codes (referred to as \textit{flags}) to determine valid combinations of candidate codes in a map. For example, a flag code of 00000 or 10000 was used to represent a one-to-one relationship. The flag code of 00000 signified the exact equivalence, whereas a flag code of 10000 represented the approximate equivalence. If the relationship were one-to-many, the third character in the flag code would be 1 (instead of 0), and the fourth and fifth characters would specify combinations of candidate codes. The fourth character enumerated the number of scenarios, while the fifth character established the order that combinations were carried out in each scenario. The data used in this paper can be obtained directly from the CMS website at \url{https://www.cms.gov/Medicare/Coding/ICD10/Archive-ICD-10-CM-ICD-10-PCS-GEMs}. The 2015 GEMs, instead of the newer GEMs, are utilized here since they were the most updated mappings available to users to prepare for the transition from ICD-9-CM to ICD-10-CM/PCS in 2015.
\subsection{Demonstration}
\begin{figure}[ht!]
	\begin{tikzpicture}
		\node [yshift = 2.5cm,right=10cm] (nodea1){(02H43JZ)};
		\node [yshift = -0.4cm] (nodea2)at (nodea1.south){(02H43KZ)};
		\node [yshift = -0.4cm] (nodea3)at (nodea2.south){(02H43MZ)};
		\node [yshift = -0.4cm] (nodea4)at (nodea3.south){(02H43KZ, 02PA0MZ)};
		\node [yshift = -0.4cm] (nodea5)at (nodea4.south){(02H43KZ, 02PA3MZ)};
		\node [yshift = -0.4cm] (nodea6)at (nodea5.south){(02H43KZ, 02PA4MZ)};
		\node [yshift = -0.4cm] (nodea7)at (nodea6.south){(02H43MZ, 02PA0MZ)};
		\node [yshift = -0.4cm] (nodea8)at (nodea7.south){(02H43MZ, 02PA3MZ)};
		\node [yshift = -0.4cm] (nodea9)at (nodea8.south){(02H43MZ, 02PA4MZ)};
		\node [right = -3cm] (nodea0)at (nodea5.west){};
		\draw [->, line width=1pt] (nodea0) -- (10.1,2.5)(nodea1);
		\draw [->, line width=1pt] (nodea0) -- (10.1,1.8)(nodea2);
		\draw [->, line width=1pt] (nodea0) -- (10.05,1.1)(nodea2);
		\draw [->, line width=1pt] (nodea0) -- (9.2,0.4)(nodea2);
		\draw [->, line width=1pt] (nodea0) -- (9.2,-0.3)(nodea2);
		\draw [->, line width=1pt] (nodea0) -- (9.2,-1)(nodea2);
		\draw [->, line width=1pt] (nodea0) -- (9.2,-1.7)(nodea2);
		\draw [->, line width=1pt] (nodea0) -- (9.2,-2.45)(nodea2);
		\draw [->, line width=1pt] (nodea0) -- (9.2,-3.15)(nodea2);	
		
		\draw [decorate,decoration={brace,amplitude=10pt,mirror,raise=4pt},xshift=9.7cm, yshift=-3.5cm]
		(3,0.2) -- (3,6.2) node [black,midway,xshift=1cm] {\footnotesize
			$v = 9$};
		\draw [decorate,decoration={brace,amplitude=10pt,mirror,raise=4pt},xshift=1.5cm, yshift=0.2cm]
		(3,-0.2) -- (3,1.5) node [black,midway,xshift=1cm] {\footnotesize
			$m_0 = 3$};
		\draw [decorate,decoration={brace,amplitude=10pt,mirror,raise=4pt},xshift=1.5cm, yshift=-1.1cm]
		(3,-0.2) -- (3,0.9) node [black,midway,xshift=1cm] {\footnotesize
			$m_1 = 2$};
		\draw [decorate,decoration={brace,amplitude=10pt,mirror,raise=4pt},xshift=1.5cm, yshift=-3cm]
		(3,-0.2) -- (3,1.5) node [black,midway,xshift=1cm] {\footnotesize
			$m_2 = 3$};
		\draw [decorate,decoration={brace,amplitude=10pt,mirror,raise=4pt},xshift=3.7cm, yshift=3.3cm]
		(-3,-1.6) -- (-3,-6.5) node [black,midway,xshift=-1cm] {\footnotesize
			$m = 8$};
		
		\node[anchor=west] (node1) {%
			\begin{tabular}{c|c|c}
				\multicolumn{3}{c}{\textbf{Map 0052}} \\
				\midrule
				\textbf{ICD-9} & \textbf{ICD-10} & \textbf{Flag} \\
				\midrule
				0052  & 02H43JZ  & 10000 \\
				0052  & 02H43KZ  & 10000 \\
				0052  & 02H43MZ  & 10000 \\
				0052  & 02H43KZ  & 10111 \\
				0052  & 02H43MZ  & 10111 \\
				0052  & 02PA0MZ  & 10112 \\
				0052  & 02PA3MZ  & 10112 \\
				0052  & 02PA4MZ  & 10112 \\
				\bottomrule
		\end{tabular}};
		
		\node[anchor=west, yshift = -7cm] (node10) {%
			\begin{tabular}{c|c|c|c|c|c|c}
				\multicolumn{7}{c}{\textbf{ICD-10 codes}} \\
				\midrule
				$\boldsymbol{a_1}$& $\boldsymbol{a_2}$ & $\boldsymbol{a_3}$& $\boldsymbol{a_4}$ & $\boldsymbol{a_5}$ & $\boldsymbol{a_6}$& $\boldsymbol{a_7}$ \\
				\midrule
				0     & 2     & H     & 4     & 3     & J     & Z \\
				0     & 2     & H     & 4     & 3     & K     & Z \\
				0     & 2     & H     & 4     & 3     & M     & Z \\
				0     & 2     & H     & 4     & 3     & K     & Z \\
				0     & 2     & H     & 4     & 3     & M     & Z \\
				0     & 2     & P     & A     & 0     & M     & Z \\
				0     & 2     & P     & A     & 3     & M     & Z \\
				0     & 2     & P     & A     & 4     & M     & Z \\
				\bottomrule
		\end{tabular}};
		
		\node [right=1cm, yshift = 2.3cm] (node20) at (node10.east){$H(a_1)= -\frac{8}{8}\log_2 \frac{8}{8} =0 $};
		
		(\node [right=1cm, yshift = 1.4cm] (node30) at (node10.east){$H(a_2)= -\frac{8}{8}\log_2 \frac{8}{8} =0$};
		
		\node [right=1cm, yshift = 0.5cm] (node40) at (node10.east){$H(a_3)= -\frac{5}{8}\log_2\frac{5}{8} -\frac{3}{8}\log_2 \frac{3}{8}  = 0.95  $};
		
		\node [right=1cm, yshift = -0.4cm] (node50) at (node10.east){$H(a_4)= -\frac{5}{8}\log_2\frac{5}{8} -\frac{3}{8}\log_2 \frac{3}{8}  = 0.95  $};
		
		\node [right=1cm, yshift = -1.3cm] (node60) at (node10.east){$H(a_5)= -\frac{6}{8}\log_2\frac{6}{8} -\frac{1}{8}\log_2 \frac{1}{8}- \frac{1}{8} \log_2 \frac{1}{8}   = 1.06  $};
		
		\node [ right=1cm, yshift = -2.2cm] (node70) at (node10.east){$H(a_6)= -\frac{5}{8}\log_2\frac{5}{8} -\frac{2}{8}\log_2 \frac{2}{8} -\frac{1}{8}\log_2 \frac{1}{8}   = 1.3 $};
		
		\node [right=1cm, yshift = -3.1cm] (node80) at (node10.east){$H(a_7)= -\frac{8}{8}\log_2 \frac{8}{8} =0 $};
		
		\node [right=0.6cm, yshift=-0.37cm] (node300) at (node10.south){};
		
		\node [rectangle, draw,xshift = 0.75cm, yshift=-1.5cm] (node400) at (node10.south){$
			H(A) \equiv \alpha = \sum_{j = 1}^{7} H(a_j) = 4.26
			$};
		\node [text centered, minimum height = 0.95 cm, right = 0.3cm, draw] (node5) at (node400.east){$H(B)\equiv \beta = \log_2 v  = 3.17$};
		
		\node [right =2.7cm, yshift =-0.2cm] (node5r) at (node5.north){Equation \ref{eq_H(B)}};
		
		\draw [->, line width=1pt] (14.45,-0.5)node[above = 3mm, right = -5.9cm] {} |- (node5);
		
		\draw [-, line width=1pt] (13.45,-0.52) -- (14.45,-0.52);
		
		\draw [->, line width=1pt,align=center] (2.6,-3.35)  -- (2.6,-3.7)(node10);
		
		\draw [->, line width=1pt,align=center] (6,-7.4)  -- (node50);
		
		\draw [-, line width=1pt] (9.6,-10.5) -- (9.6,-11);
		\draw [-, line width=1pt] (3.75,-11)node[above = 0.2cm, right =2cm] {Equation \ref{eq_H(A)}} -- (9.6,-11);
		\draw [->, line width=1pt] (node300) -- (node400);

		\node [yshift = -1.1cm] (noden1) at (node400.south){$Z(\alpha) = \frac{4.26 - \bar{\alpha}}{v(\alpha)} = 0.572$};
		
		\node [yshift = -1.1cm] (noden2) at (node5.south){$Z(\beta) = \frac{3.17- \bar{\beta}}{v(\beta)} = 0.212$};
		
		\draw [->, line width=1pt] (node400)node[above = -0.8cm, right =0cm] {Equation \ref{eq_normalize}}  -- (noden1);
		\draw [->, line width=1pt] (node5)node[above = -0.8cm, right =0cm] {Equation \ref{eq_normalize2}} -- (noden2);
		
	\end{tikzpicture}
	\caption{This figure depicts how to apply \textbf{Steps 1-3} of Algorithm \ref{algorithm_shannon} on map 0052. No empirical data were available to implement \textbf{Step 4} of this algorithm. $H(A)$ is computed per Equation \ref{eq_H(A)}. Equation \ref{eq_v} is used to determine the number of valid representations $v$ and $H(B)$ is calculated per Equation \ref{eq_H(B)}. The normalization of  $H(A)$ and $H(B)$ follow Equations \ref{eq_normalize} and \ref{eq_normalize2}, respectively. The $UR$ measure (proposed in Chen et al. \cite{chen2018leveraging}) of this map is obtained by $\log_2 (m)=\log_2 (8) = 3$. As compared to $H(B)$, the UR measure slightly underestimates the complexity of map 0052. While no actual probabilities were available for \textbf{Step 4}, it still can be speculated that if the probability of implanting or replacing any electrodes in the ventricular coronary venous system were zero for a given medical facility, both the $Z(\alpha)$ and $Z(\beta)$ measures would be 0.572*0 = 0.212*0 = 0. The implementation of \textbf{Step 5} of this algorithm is discussed in Section \ref{sec_results}.}
	\label{fig_example}  
\end{figure}
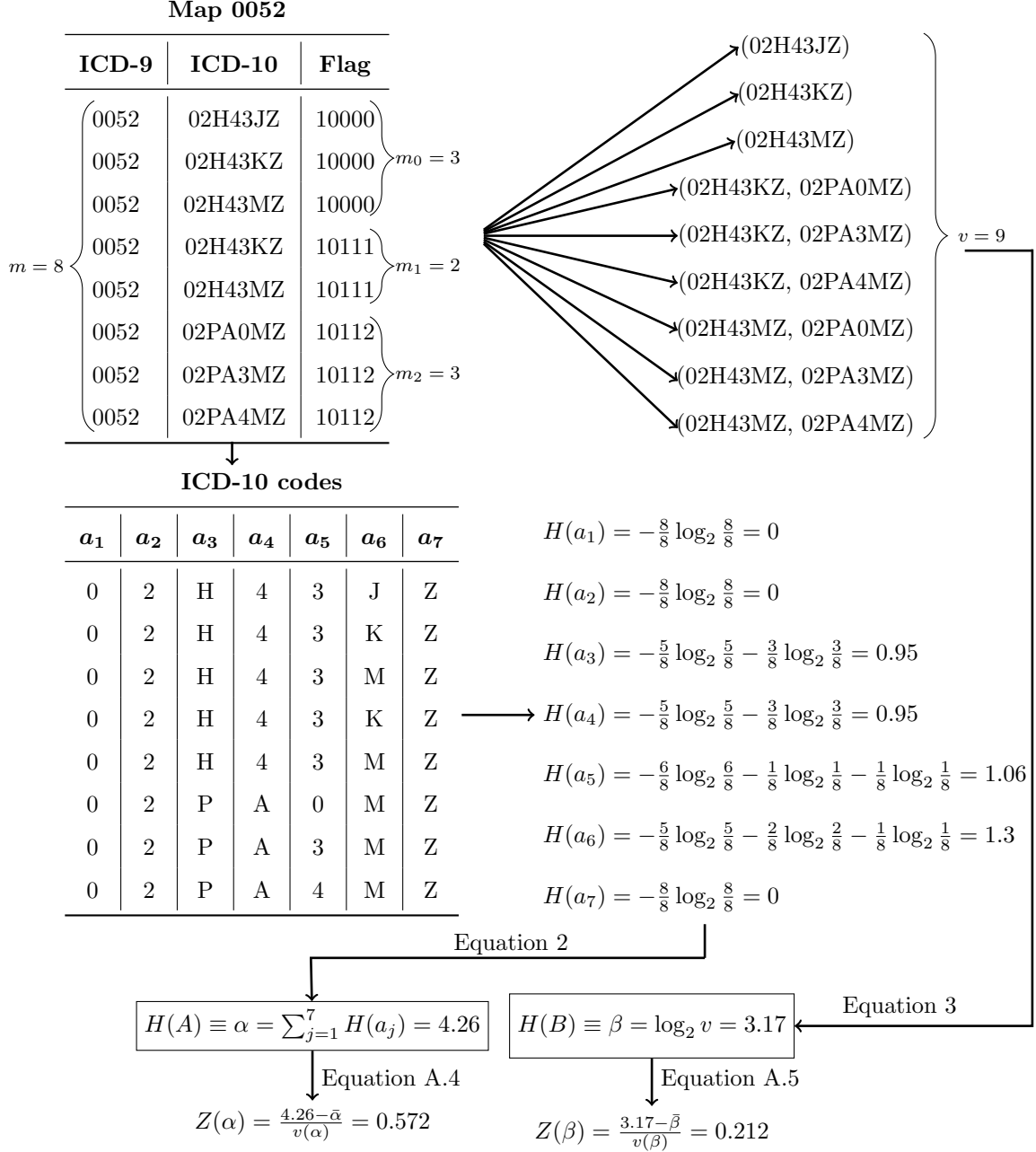

\ref{appendix_python} demonstrates a Python code to implement Algorithm \ref{algorithm_shannon}. Figure \ref{fig_example} exhibits the application of this algorithm on map 0052. This map relates to an ICD-9-CM Vol.3 code of 00.52 for the \textit{implantation or replacement of transvenous electrode into left ventricular coronary venous system}. 

\section{RESULTS} \label{sec_results}
\begin{figure}[ht!]
	\centering  
	\begin{subfigure}[b]{0.57\textwidth}
		\centering
		\tcbox[size=fbox,on line]{\includegraphics[trim = 0mm 0mm 0mm 0mm, clip, width=10.1cm]{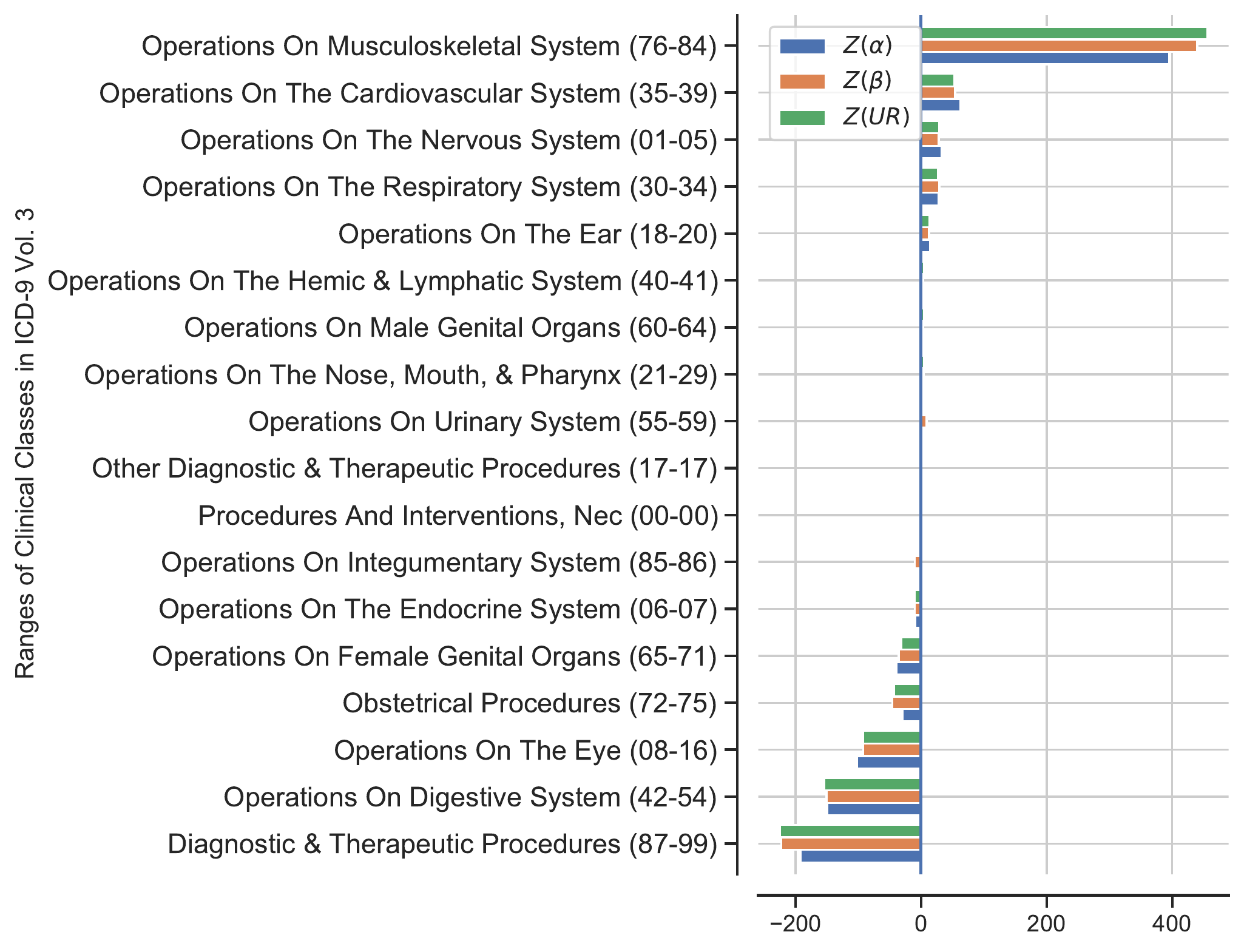}}
		\caption{Forward entropic measures from ICD-9-CM Vol.3 to ICD-10-PCS }
		\label{fig_9-PCS-bar}
	\end{subfigure}%
	\begin{subfigure}[b]{0.5\textwidth}
		\centering
		\tcbox[size=fbox,on line]{\includegraphics[trim = 0mm 0mm 0mm 0mm, clip, width=5.08cm]{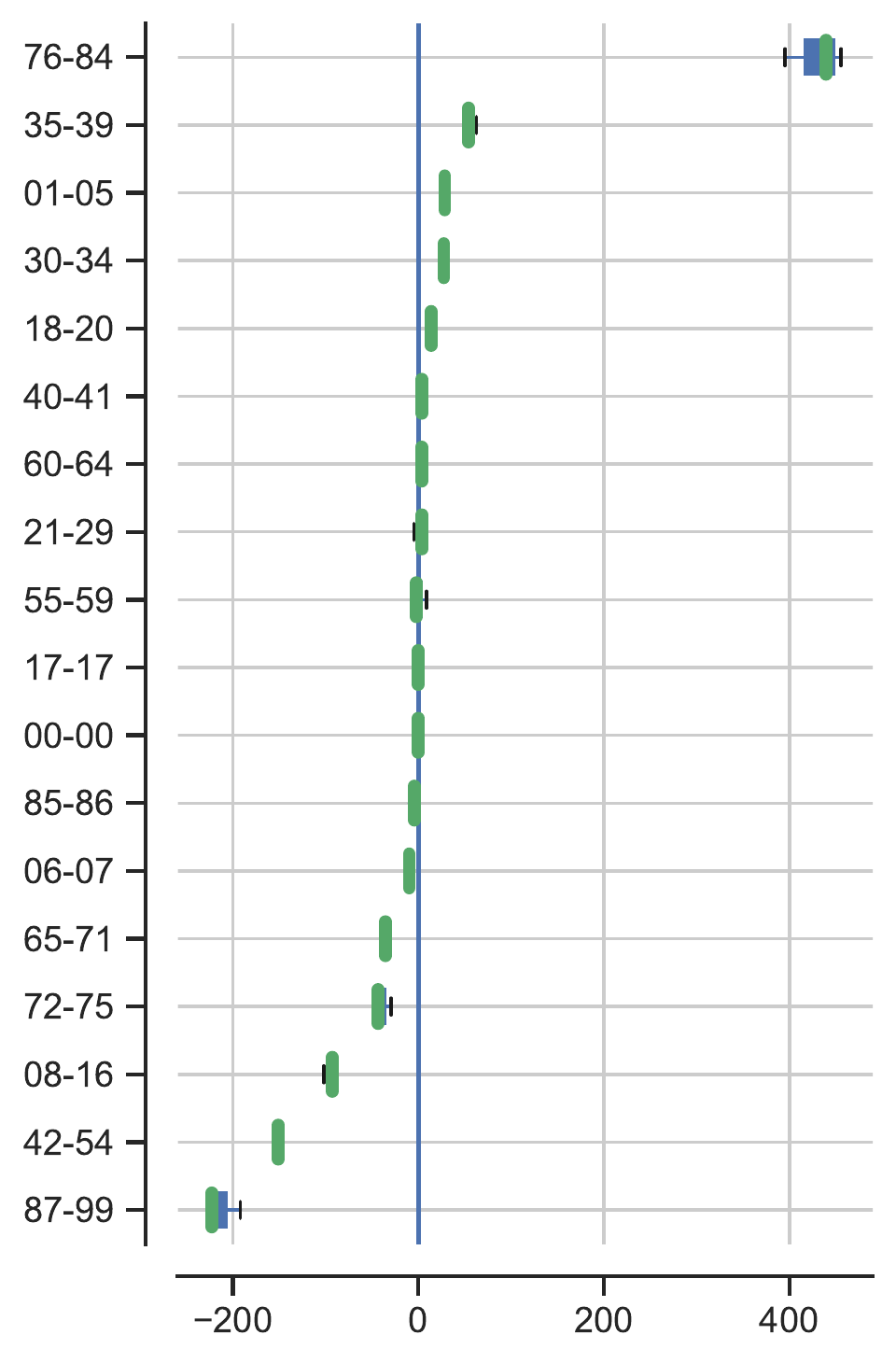}}
		\caption{Box plots of entropic measures in \textbf{(a)}}
		\label{fig_9-PCS-box}
	\end{subfigure}  
	\smallskip
	
	\begin{subfigure}[b]{0.57\textwidth}
		\centering
		\tcbox[size=fbox,on line]{\includegraphics[trim = 0mm 0mm 0mm 0mm, clip, width=10.1cm]{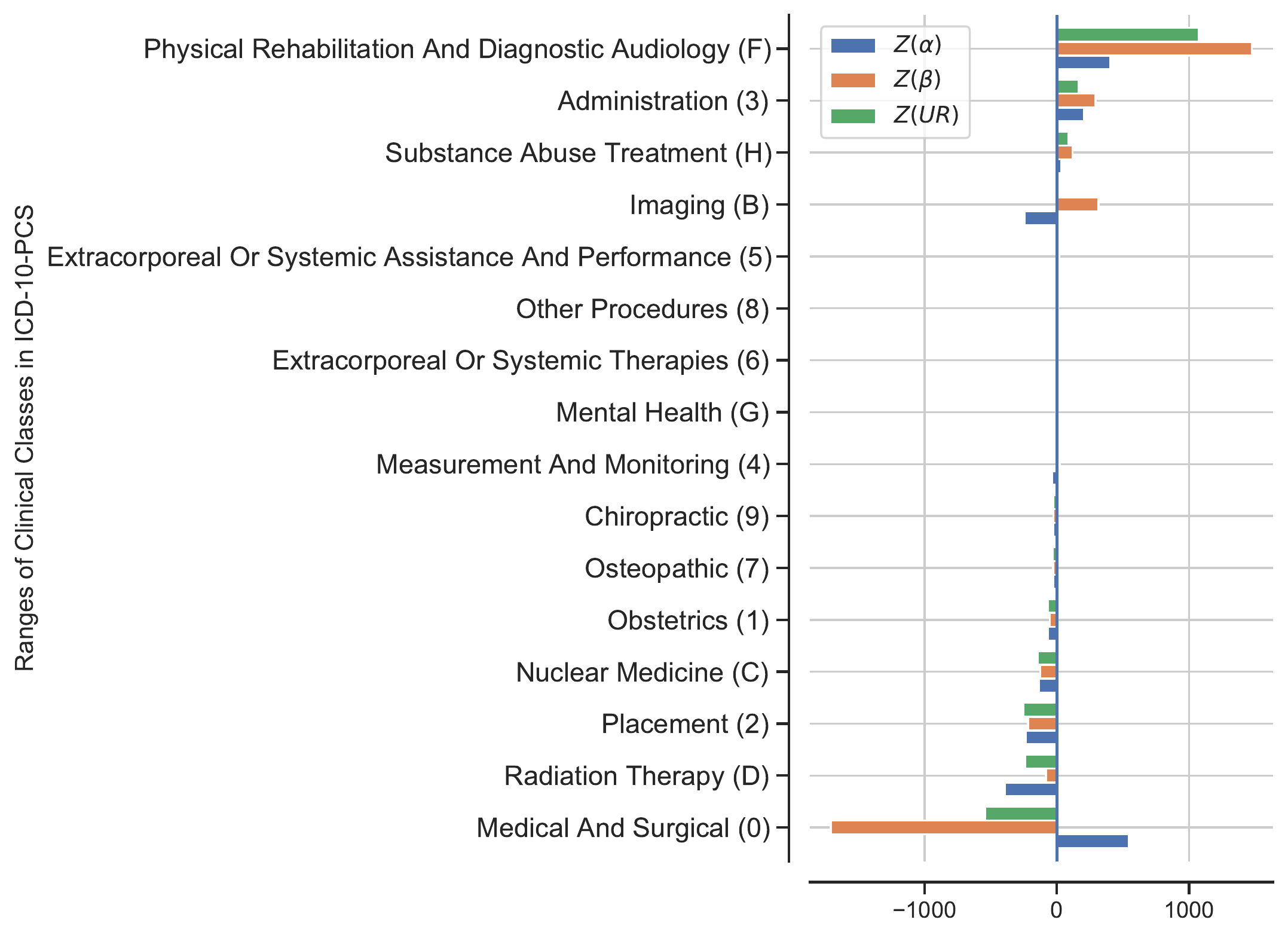}}
		\caption{Backward entropic measures from ICD-10-PCS to ICD-9-CM Vol.3}
		\label{fig_PCS-9-bar}
	\end{subfigure}%
	\begin{subfigure}[b]{0.5\textwidth}
		\centering
		\tcbox[size=fbox,on line]{\includegraphics[trim = 0mm 0mm 0mm 0mm, clip, width=4.9cm]{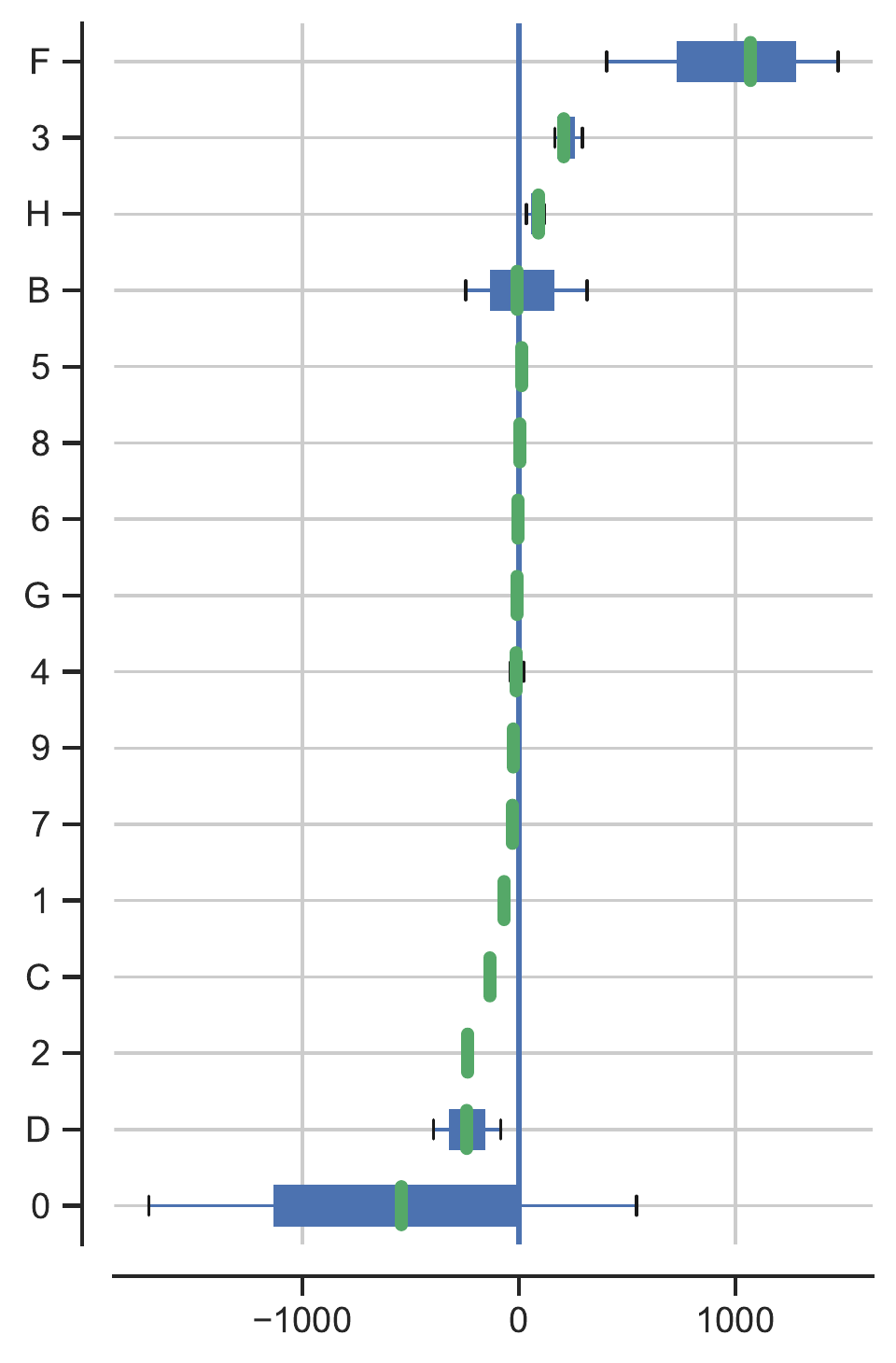}}
		\caption{Box plots of entropic measures in \textbf{(c)}}
		\label{fig_PCS-9-box}
	\end{subfigure}  
	\caption{Forward and backward entropic measures between the procedure codes of ICD-9-CM Vol.3 and ICD-10-PCS. The x-axes represent the sum of $Z(\alpha)$, $Z(\beta)$, and $Z(UR)$ entropic measures. Sub-figures \ref{fig_9-PCS-bar} and \ref{fig_PCS-9-bar} show clustered bar plots of the indicated clinical classes arranged from the least to the most sum of entropic measures. Negative values signify no information gained or lost information (on average) from the source system to the target system.  Positive values suggest gained information. Sub-figures \ref{fig_9-PCS-box} and \ref{fig_PCS-9-box} display related box plots that may help visually assess the variation in the entropic measures in each clinical class. The wider the box, the more the interquartile range, thus the more variability in the measures. The tighter the box and whiskers, the more the measures agree.} \label{fig_procedure}  
\end{figure}

\begin{figure}[ht!]
	\centering  
	\begin{subfigure}[b]{0.57\textwidth}
		\centering
		\tcbox[size=fbox,on line]{\includegraphics[trim = 0mm 0mm 0mm 0mm, clip, width=10.6cm]{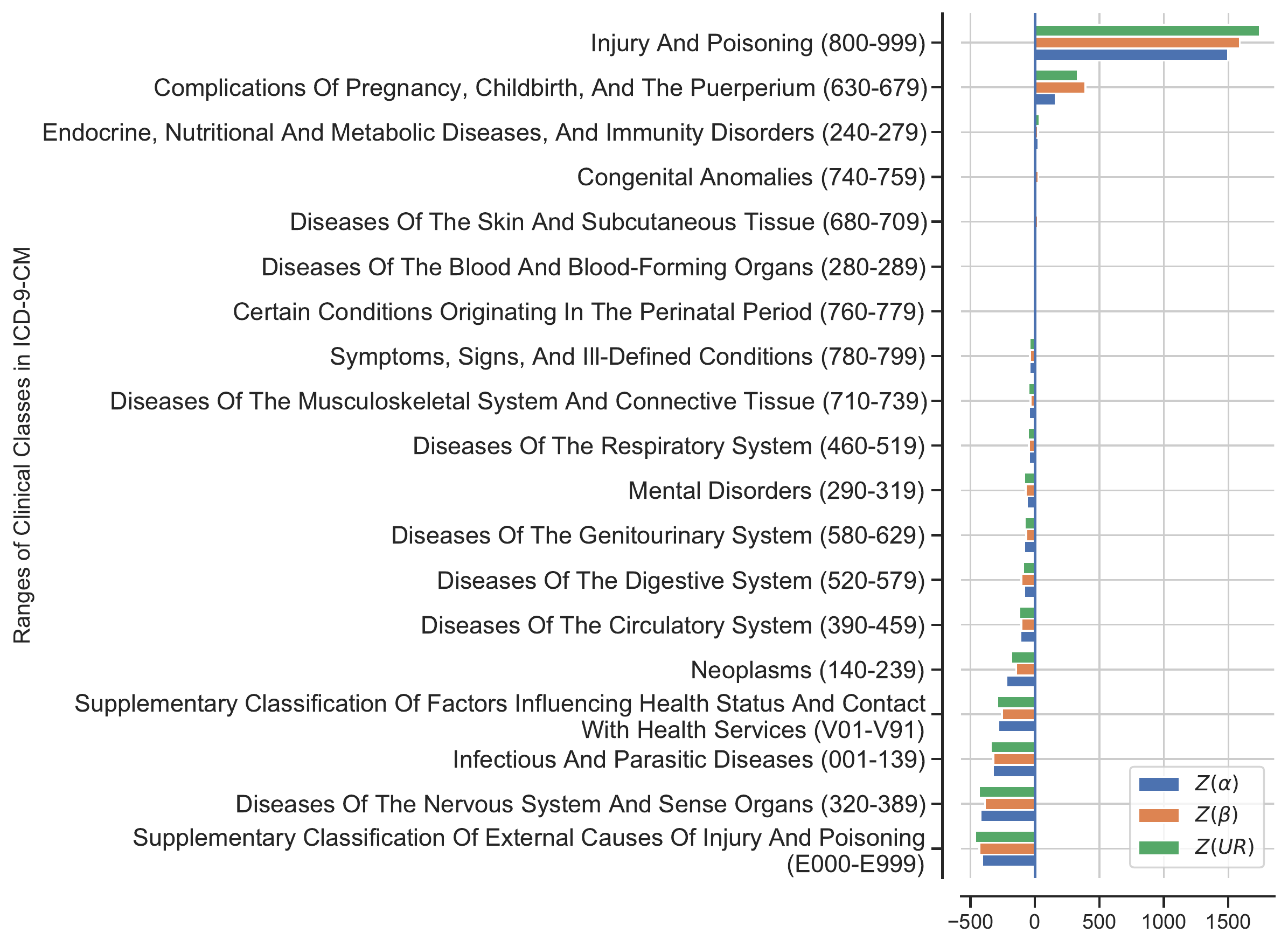}}
		\caption{Forward entropic measures from ICD-9-CM to ICD-10-CM}
		\label{fig_CM9-10-bar}
	\end{subfigure}%
	\begin{subfigure}[b]{0.5\textwidth}
		\centering
		\tcbox[size=fbox,on line]{\includegraphics[trim = 0mm 0mm 0mm 0mm, clip, width=4.77cm]{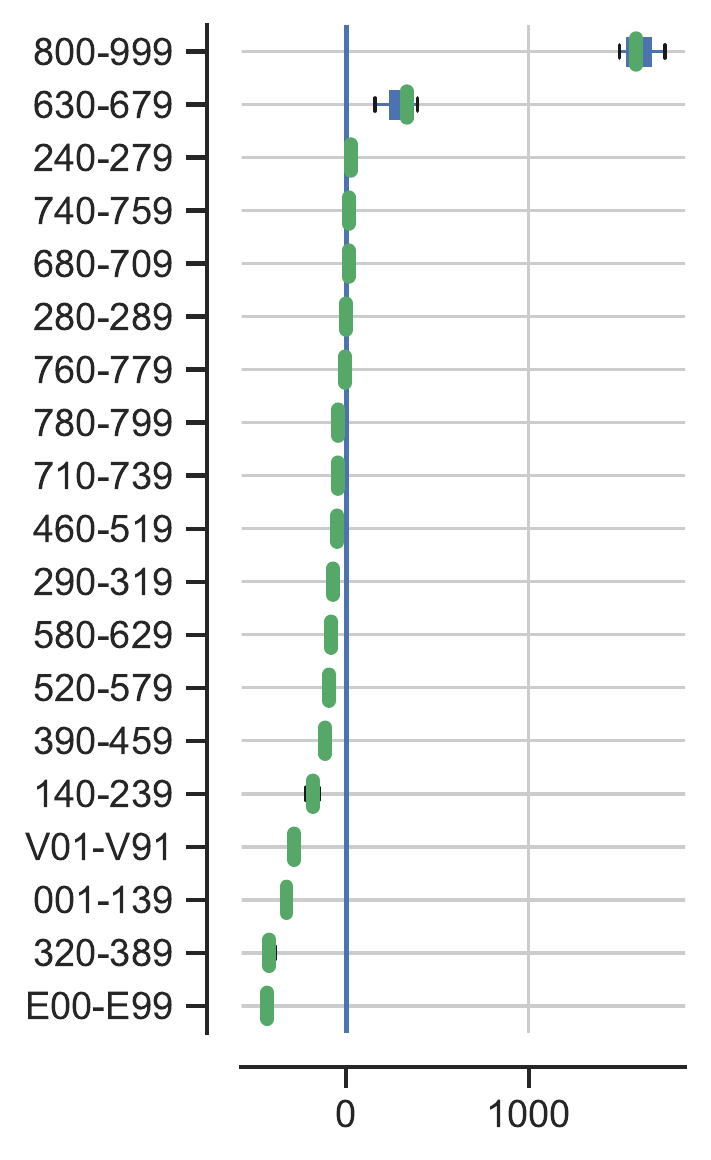}}
		\caption{Box plots of entropic measures in \textbf{(a)}}
		\label{fig_CM9-10-box}
	\end{subfigure}  
	\smallskip

	\begin{subfigure}[b]{0.57\textwidth}
		\centering
		\tcbox[size=fbox,on line]{\includegraphics[trim = 0mm 0mm 0mm 0mm, clip, width=10.69cm]{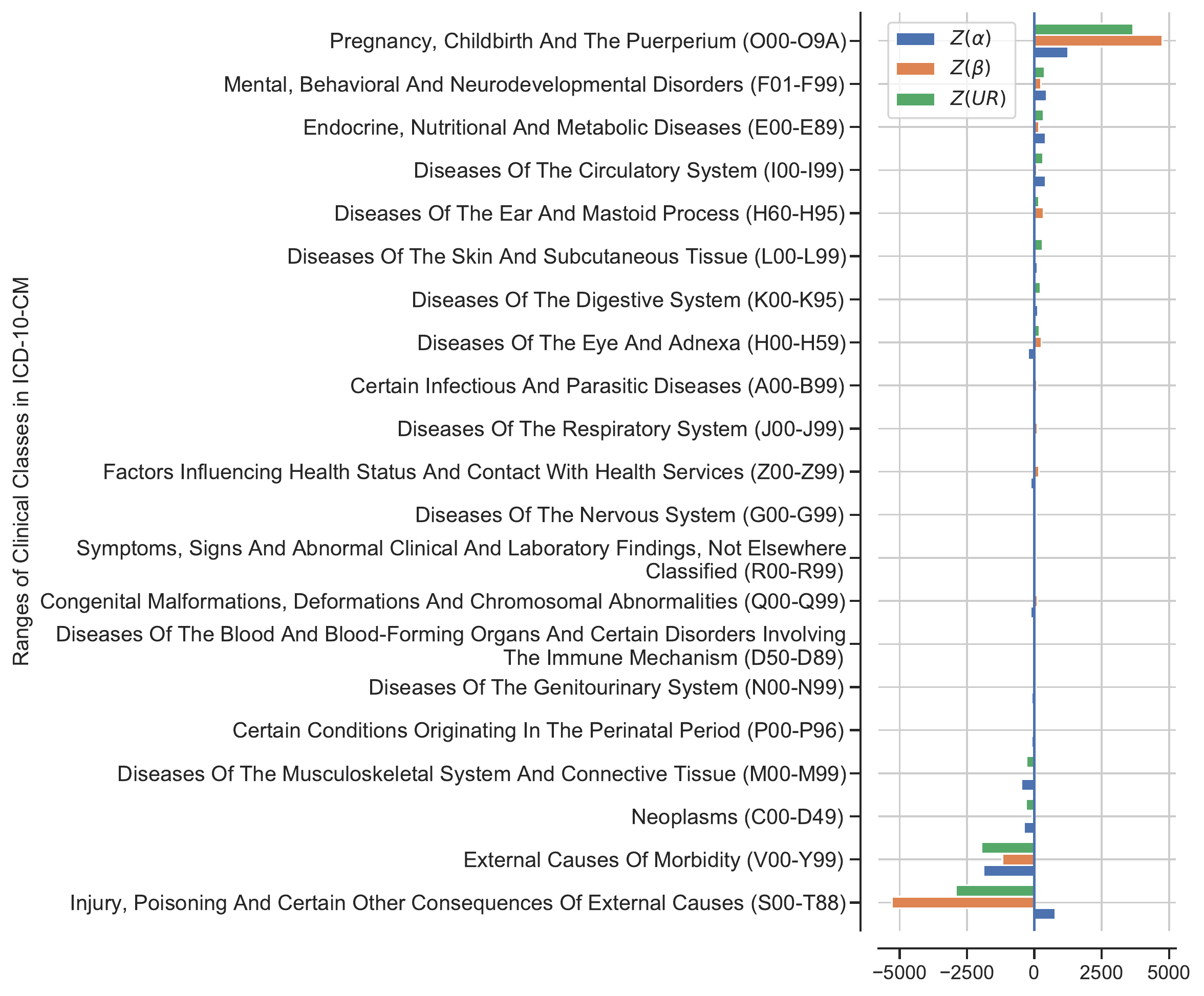}}
		\caption{Backward entropic measures from ICD-10-CM to ICD-10-CM }
		\label{fig_cm10-9-bar}
	\end{subfigure}%
	\begin{subfigure}[b]{0.5\textwidth}
		\centering
		\tcbox[size=fbox,on line]{\includegraphics[trim = 0mm 0mm 0mm 0mm, clip, width=4.58cm]{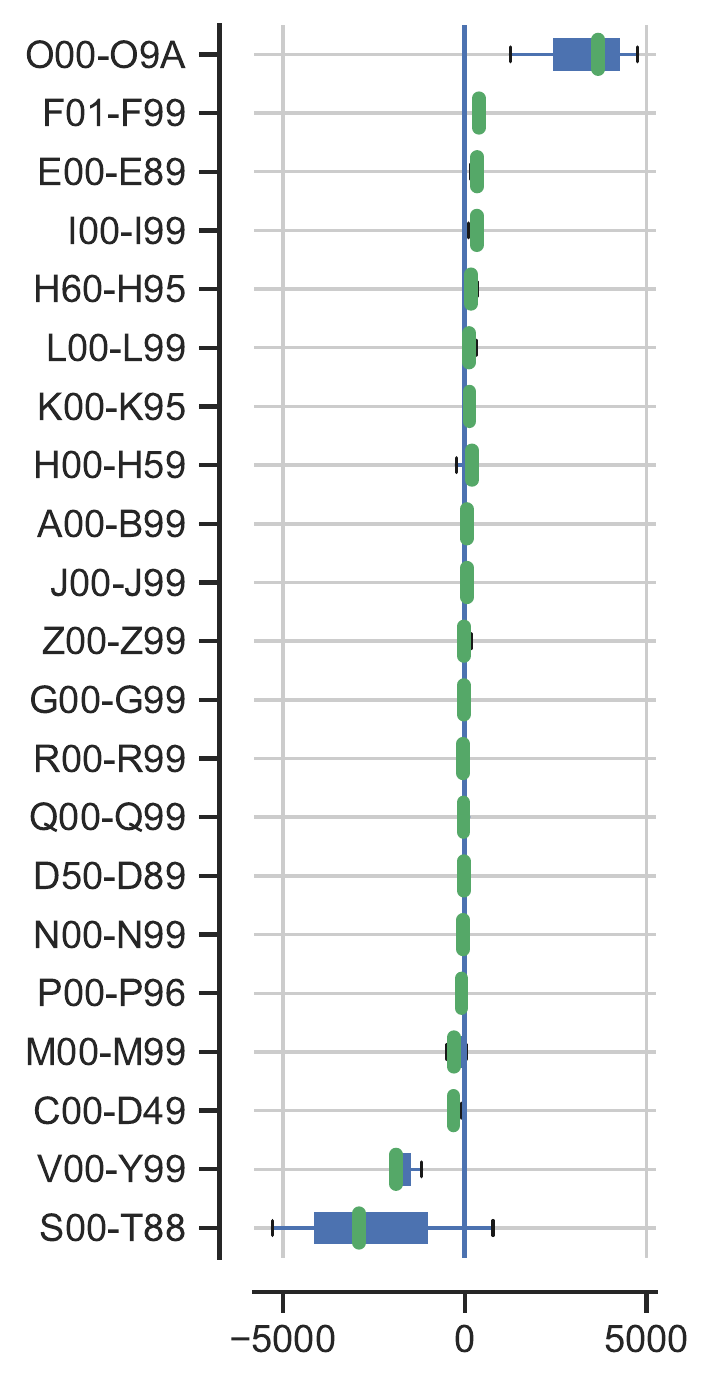}}
		\caption{Box plots of entropic measures in \textbf{(c)}}
		\label{fig_cm10-9-box}
	\end{subfigure}  
	\caption{Forward and backward entropic measures between the diagnosis codes of ICD-9-CM and ICD-10-CM. The x-axes represent the sum of $Z(\alpha)$, $Z(\beta)$, and $Z(UR)$ entropic measures. Sub-figures \ref{fig_CM9-10-bar} and \ref{fig_cm10-9-bar} show clustered bar plots of the indicated clinical classes arranged from the least to the most sum of entropic measures. Negative values signify no information gained or lost information (on average) from the source system to the target system.  Positive values suggest gained information. Sub-figures \ref{fig_CM9-10-box} and \ref{fig_cm10-9-box} display related box plots that may help visually assess the variation in the entropic measures in each clinical class. The wider the box, the more the interquartile range, thus the more variability in the measures. The tighter the box and whiskers, the more the measures agree.} \label{fig_diagnosis}  
\end{figure}

\begin{figure}[h!]
	\centering  
	\begin{subfigure}[b]{0.5\textwidth}
		\centering
		\tcbox[size=fbox,on line]{\includegraphics[trim = 0mm 0mm 0mm 0mm, clip, width=7.8cm]{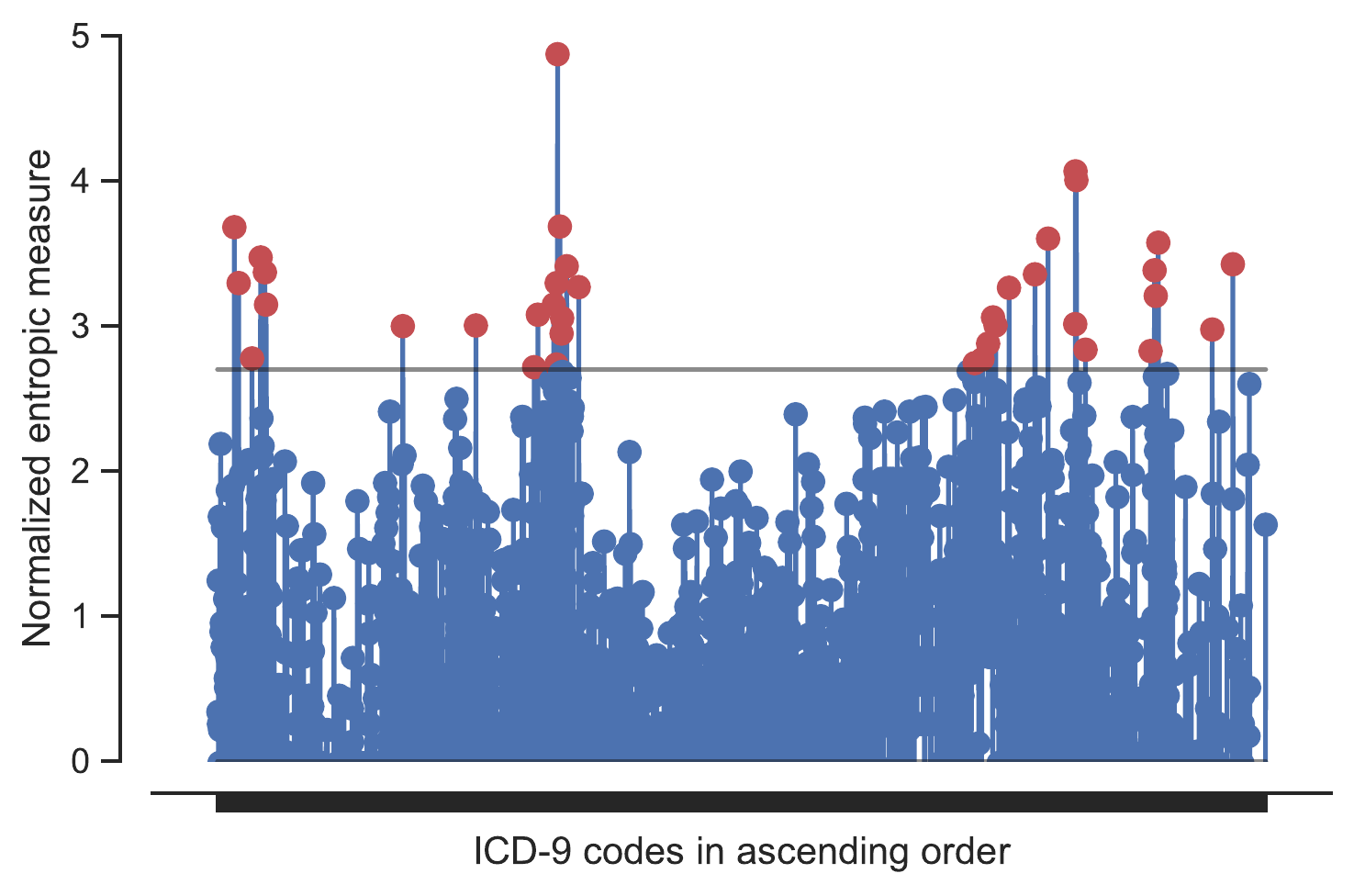}}
		\caption{$Z(\alpha)$ outlier maps are in red color}
		\label{fig_control1}
	\end{subfigure}%
	\begin{subfigure}[b]{0.5\textwidth}
		\centering
		\tcbox[size=fbox,on line]{\includegraphics[trim = 0mm 0mm 0mm 0mm, clip, width=7.8cm]{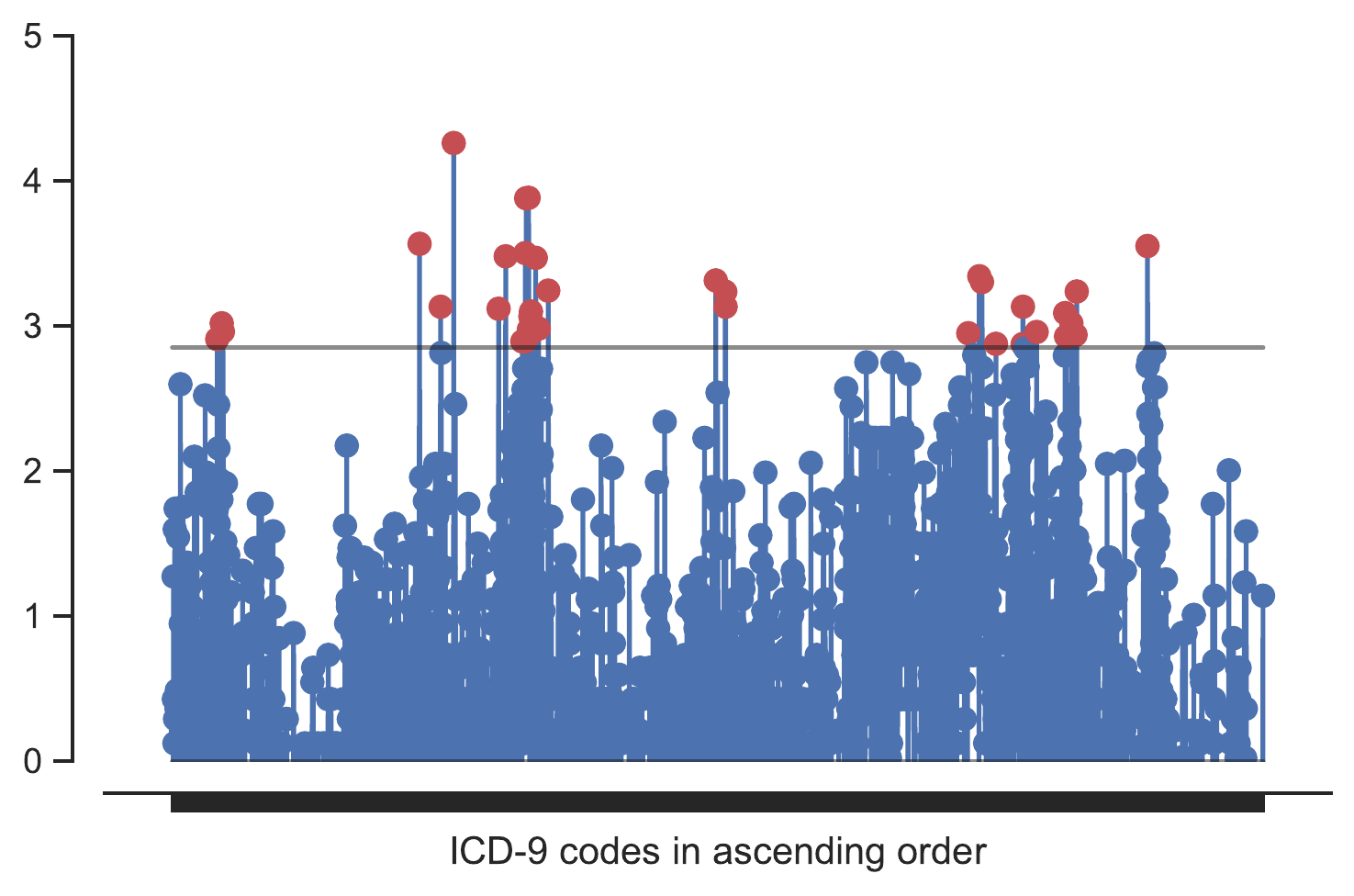}}
		\caption{$Z(\beta)$ outlier maps are in red color}
		\label{fig_control2}
	\end{subfigure}  
	\\
	\begin{subfigure}[b]{0.5\textwidth}
		\centering
		\tcbox[size=fbox,on line]{\includegraphics[trim = 0mm 50mm 0mm 50mm, clip, width=7.8cm]{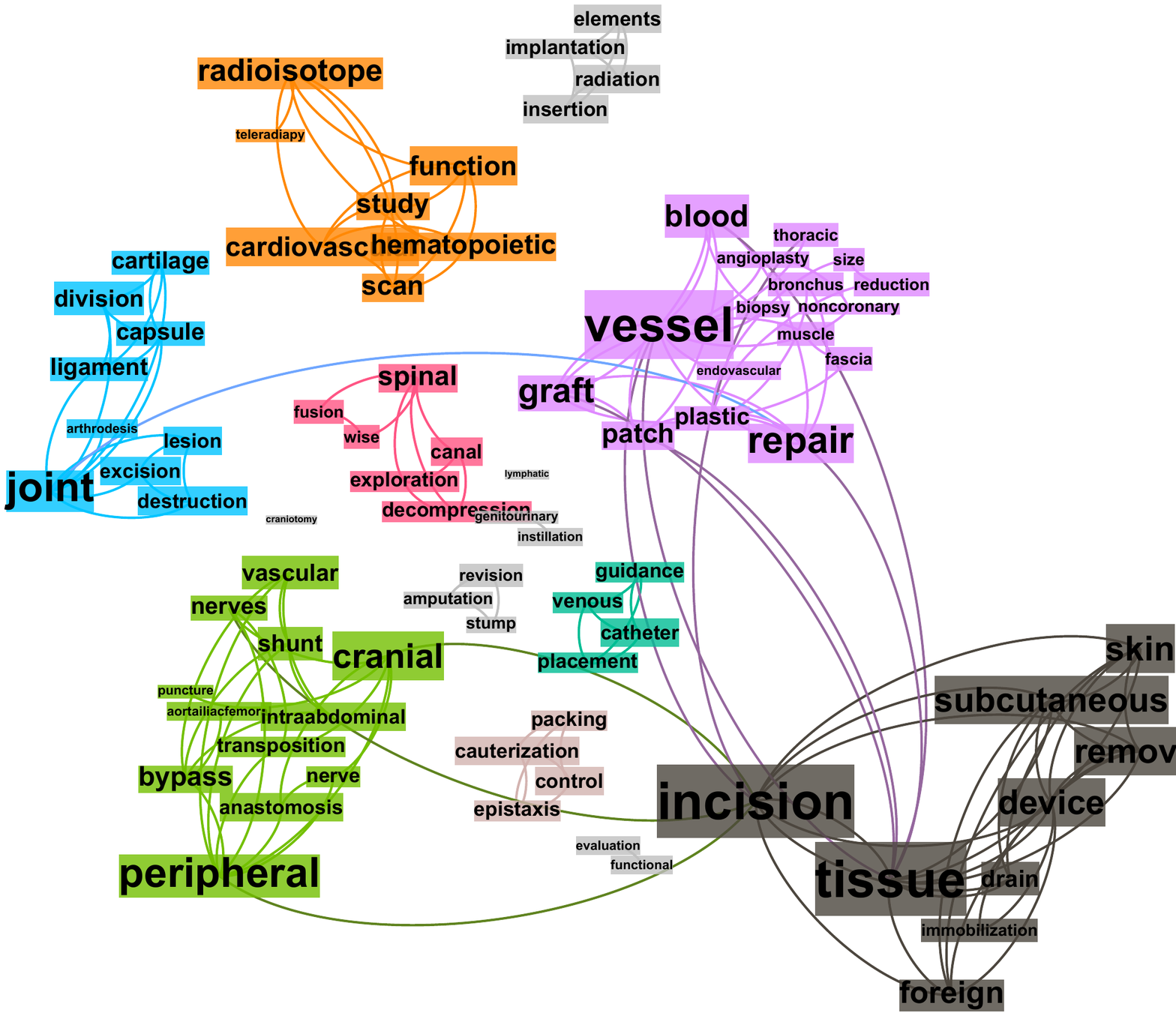}}
		\caption{Network of words from the descriptions of $Z(\alpha)$ outliers}
		\label{fig_df31}
	\end{subfigure}%
	\begin{subfigure}[b]{0.5\textwidth}
		\centering
		\tcbox[size=fbox,on line]{\includegraphics[trim = 0mm 50mm 0mm 50mm, clip, width=7.8cm]{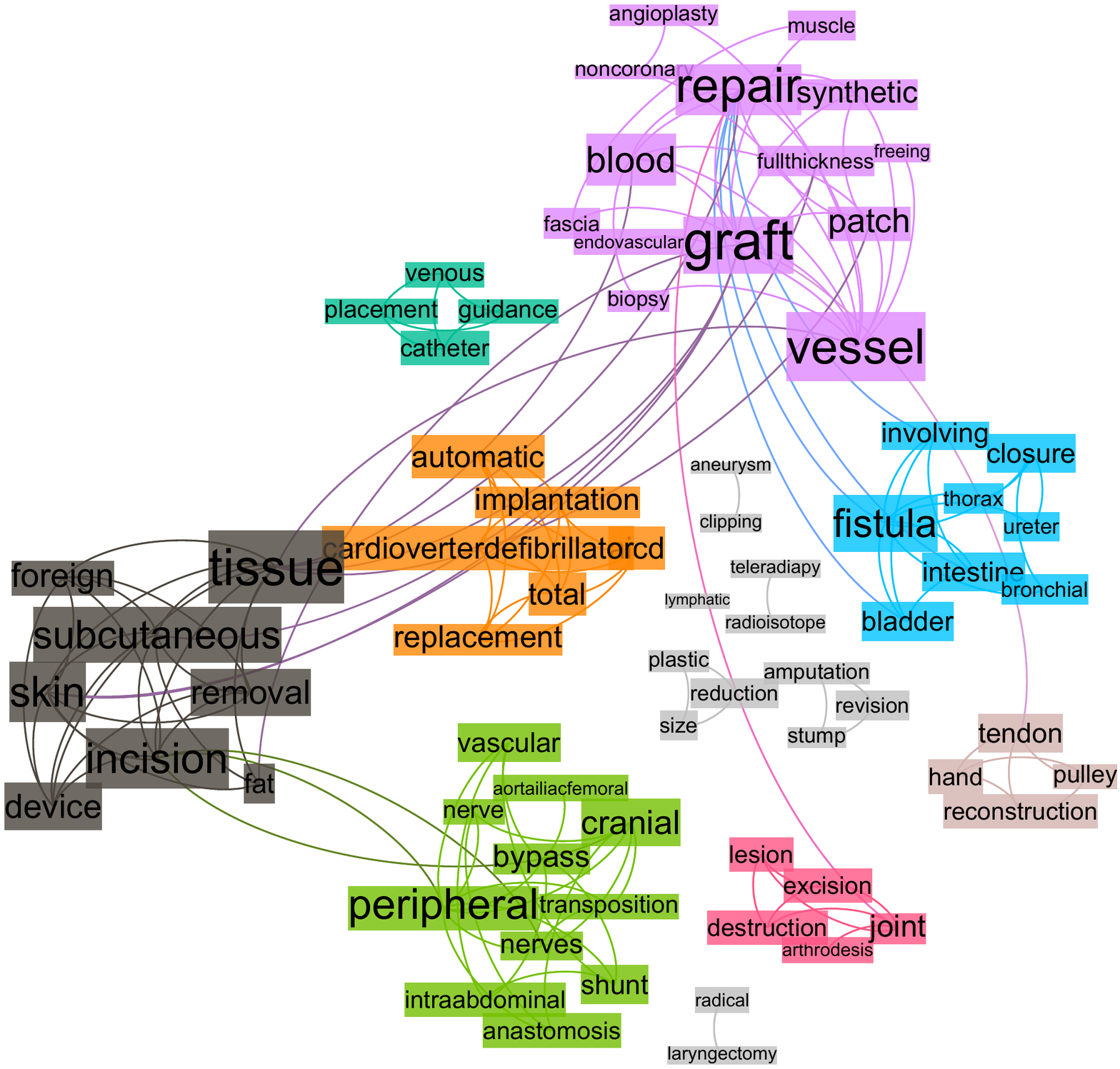}}
		\caption{Network of words from the descriptions of $Z(\alpha)$ outliers}
		\label{fig_df32}
	\end{subfigure} 
	\\
	\begin{subfigure}[b]{0.5\textwidth}
		\centering
		\tcbox[size=fbox,on line]{\includegraphics[trim = 0mm 0mm 0mm 0mm, clip, width=7.8cm]{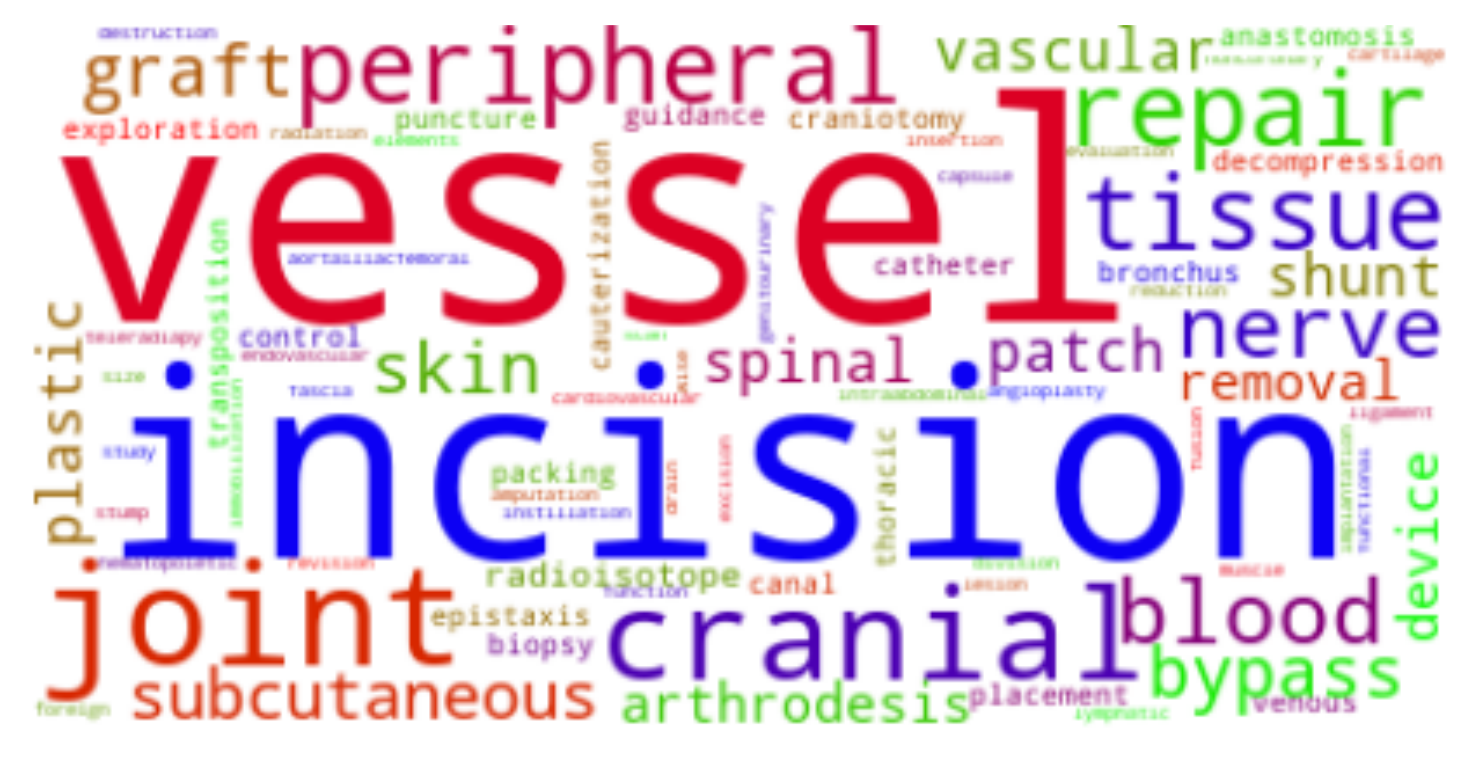}}
		\caption{Word cloud based on the descriptions of $Z(\alpha)$ outliers}
		\label{fig_df31Cloud}
	\end{subfigure}%
	\begin{subfigure}[b]{0.5\textwidth}
		\centering
		\tcbox[size=fbox,on line]{\includegraphics[trim = 0mm 0mm 0mm 0mm, clip, width=7.8cm]{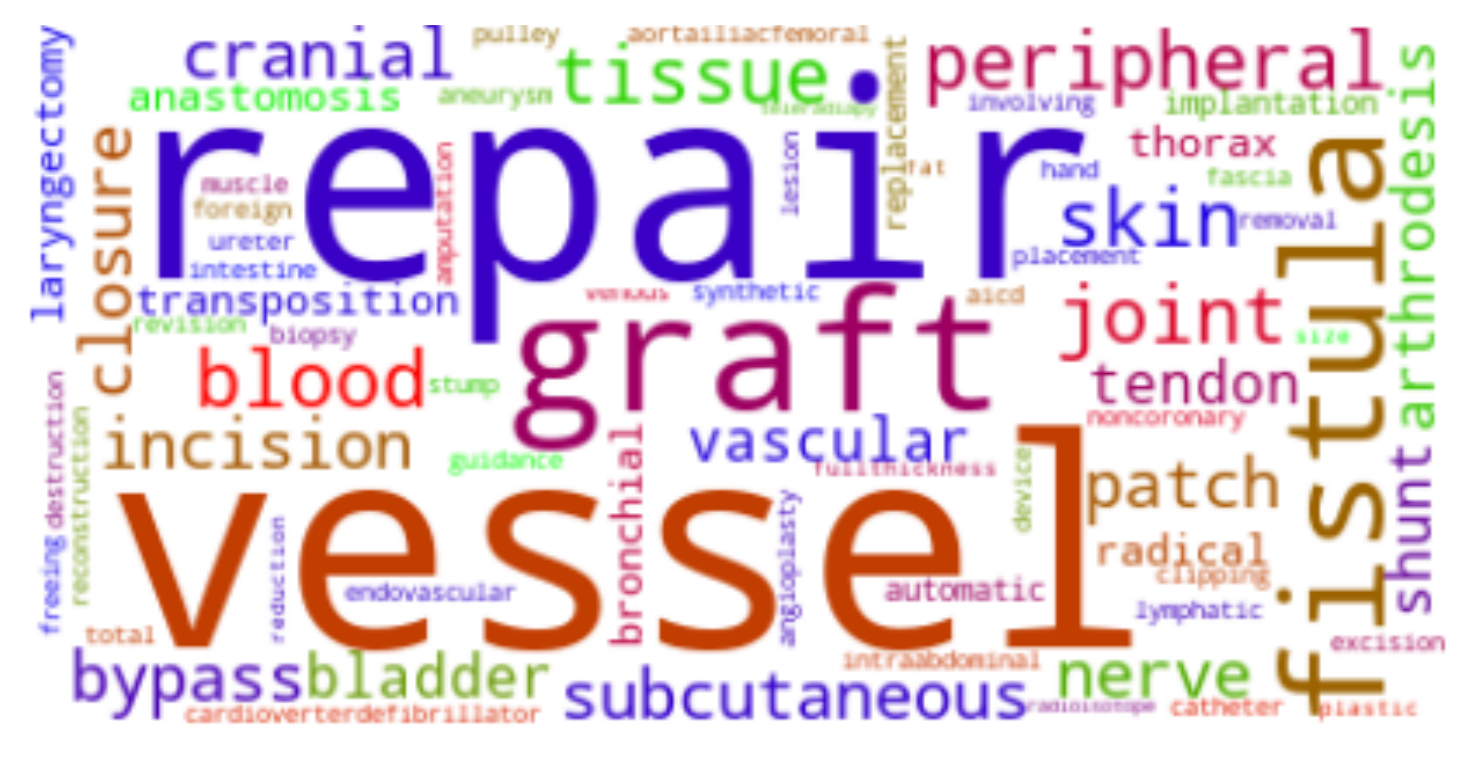}}
		\caption{Word cloud based on the descriptions of $Z(\beta)$ outliers}
		\label{fig_df32Cloud}
	\end{subfigure}  
	\caption{An example of outlier and pattern analysis based on forward mapping from ICD-9-CM Vol.3  and ICD-10-PCS. Sub-figures \ref{fig_control1} and \ref{fig_control2} show red dots for outlier maps where $Z(\alpha)$ and $Z(\beta)$ scores are greater than the chosen threshold. For illustration purposes, the threshold were determined (2.7 for  $Z(\alpha)$ and 2.85 for $Z(\beta)$, so that only the top 1\% of the cases are isolated. The maps that were isolated are shown in \ref{appendix_za} and \ref{appendix_zb}. Sub-figure \ref{fig_df31} illustrates a network of words in the descriptions of $Z(\alpha)$ outlier maps while Sub-figure \ref{fig_df32} portrays a similar network for $Z(\beta)$ outlier maps. The corresponding word clouds are respectively shown in Sub-figures \ref{fig_df31Cloud} and \ref{fig_df32Cloud}.}
	\label{fig_controls}  
\end{figure}

Algorithm \ref{algorithm_shannon} was applied to both forward and backward GEMs between ICD-9-CM and ICD-10-CM/PCS. Tables \ref{tab_procedure} and \ref{tab_diagnosis} display the corresponding descriptive statistics for $H(A)$, $H(B)$, and $UR$ entropic measures. Codes without match in the target system were excluded from these statistics. For comparison purposes, the normalization of the $UR$  measure \cite{chen2018leveraging} is symbolized as $Z(UR)$. To implement \textbf{Step 5} of Algorithm \ref{algorithm_shannon}, clinical concepts were ranked by their entropic measures. Figures \ref{fig_procedure} and \ref{fig_diagnosis} show ranked clinical classes from the least to the most sum of $Z(\alpha)$, $Z(\beta)$, and $Z(UR)$ measures. The classes in these figures were also ranked separately using each entropic measure. As expected, the resulting rankings based on $Z(\alpha)$, $Z(\beta)$, and $Z(UR)$ measures were not always consistent. To assess how much the rankings of these entropic measures agreed, the Kendall tau correlation coefficients were assessed, and the results are presented in Table \ref{tab_corr}. The closer to 1 the Kendall tau value (the greener the color), the more the given entropic measures agreed. An alternative approach to implementing \textbf{Step 5} of Algorithm \ref{algorithm_shannon} is performing outlier and pattern analysis and then segregate concepts that should receive more attention during the transition. An example of how such an analysis may be conducted is shown in Figure \ref{fig_controls}. To extract thematic descriptions of the outlier maps, network analysis techniques suggested in Niyirora and Aragones \cite{niyirora2019network} were applied after removing stopwords \citep{wilbur1992automatic} and residual words (e.g., other, unspecified, etc.).  Communities of words in Sub-figures \ref{fig_df31} and \ref{fig_df32} (distinguished by different colors) were isolated using the modularity algorithm in Gephi \citep{bastian2009gephi}. To gauge the frequency (or significance) of words in the outlier maps, a word cloud analysis was undertaken, where the bigger the word meant, the more significant the word (see Sub-figures \ref{fig_df31Cloud} and \ref{fig_df32Cloud}).
\begin{table}[h!]
	\centering
	\caption{Descriptive statistics of the $H(A)$, $H(B)$, and $UR$ entropic measures between ICD-9-CM Vol.3 and ICD-10-PCS}
	\scalebox{0.8}{
	\begin{tabular}{c|p{2cm}p{2cm}p{2cm}|p{2cm}p{2cm}p{2cm}}
		\toprule
		\multicolumn{1}{c}{\multirow{2}[2]{*}{}} & \multicolumn{3}{c|}{\textbf{Forward mapping}} & \multicolumn{3}{c}{\textbf{Backward mapping}} \\
		\multicolumn{1}{c}{} & \multicolumn{3}{c|}{\textbf{ From ICD-9-CM Vol.3  to ICD-10-PCS}} & \multicolumn{3}{c}{\textbf{From ICD-10-PCS to ICD-9-CM Vol.3}} \\
		\cmidrule{2-7}    \multicolumn{1}{c}{} & \textbf{H(A)} & \textbf{H(B)} & \textbf{UR} & \textbf{H(A)} & \textbf{H(B)} & \textbf{UR} \\
		\midrule
		\textbf{count} & 3672  & 3672  & 3672  & 71924 & 71924 & 71924 \\
		\textbf{mean} & 2.76  & 3.03  & 2.74  & 0.09  & 0.25  & 0.13 \\
		\textbf{std} & 1.92  & 2.16  & 1.85  & 0.32  & 0.84  & 0.40 \\
		\textbf{min} & 0.00  & 0.00  & 0.00  & 0.00  & 0.00  & 0.00 \\
		\textbf{25\%} & 1.00  & 1.58  & 1.44  & 0.00  & 0.00  & 0.00 \\
		\textbf{50\%} & 2.58  & 2.58  & 2.58  & 0.00  & 0.00  & 0.00 \\
		\textbf{75\%} & 4.00  & 4.39  & 3.91  & 0.00  & 0.00  & 0.00 \\
		\textbf{max} & 10.95 & 13.53 & 10.22 & 3.46  & 7.50  & 3.46 \\
	\end{tabular}%
	}
	\label{tab_procedure}%
\end{table}%

\begin{table}[h!]
	\centering
	\caption{Descriptive statistics of the $H(A)$, $H(B)$, and $UR$ entropic measures between ICD-9-CM and ICD-10-CM}
	\scalebox{0.8}{
	\begin{tabular}{c|p{2cm}p{2cm}p{2cm}|p{2cm}p{2cm}p{2cm}}
		\toprule
		\multicolumn{1}{c}{\multirow{2}[2]{*}{}} & \multicolumn{3}{c|}{\textbf{Forward mapping}} & \multicolumn{3}{c}{\textbf{Backward mapping}} \\
		\multicolumn{1}{c}{} & \multicolumn{3}{c|}{\textbf{ From ICD-9-CM to ICD-10-CM}} & \multicolumn{3}{c}{\textbf{From ICD-10-CM to ICD-9-CM}} \\
		\cmidrule{2-7}    \multicolumn{1}{c}{} & \textbf{$H(A)$} & \textbf{$H(B)$} & \textbf{$UR$} & \textbf{$H(A)$} & \textbf{$H(B)$} & \textbf{$UR$} \\
		\midrule
		\textbf{count} & 14567 & 14567 & 14567 & 69823 & 69823 & 69823 \\
		\textbf{mean} & 0.52  & 0.30  & 0.33  & 0.30  & 0.07  & 0.12 \\
		\textbf{std} & 1.26  & 0.69  & 0.72  & 1.01  & 0.28  & 0.36 \\
		\textbf{min} & 0.00  & 0.00  & 0.00  & 0.00  & 0.00  & 0.00 \\
		\textbf{25\%} & 0.00  & 0.00  & 0.00  & 0.00  & 0.00  & 0.00 \\
		\textbf{50\%} & 0.00  & 0.00  & 0.00  & 0.00  & 0.00  & 0.00 \\
		\textbf{75\%} & 0.00  & 0.00  & 0.00  & 0.00  & 0.00  & 0.00 \\
		\textbf{max} & 13.10 & 9.06  & 9.06  & 7.31  & 3.58  & 3.58 \\
	\end{tabular}%
	}
	\label{tab_diagnosis}%
\end{table}%

\begin{table}[h!]
	\centering
	\caption{Kendall tau correlation among the rankings of clinical classes using the normalized entropic measures ($Z(\alpha)$, $Z(\beta)$, and $Z(UR)$). The symbol $<>$ is used to signify mapping between the indicated medical coding systems.}
	\scalebox{0.8}{
		\begin{tabular}{M{0.05cm}M{0.05cm}M{0.05cm}|c|M{1.5cm}M{1.5cm}M{1.5cm}|p{1.5cm}p{1.5cm}p{1.5cm}}
			&       & \multicolumn{1}{r}{} & \multicolumn{1}{r}{} & \multicolumn{3}{c|}{\textbf{Forward mapping}} & \multicolumn{3}{c}{\textbf{Backward mapping}} \\
			\cmidrule{5-10}          &       & \multicolumn{1}{r}{} & \multicolumn{1}{r}{} & $Z(\alpha)$  & $Z(\beta)$  & $Z(UR)$   &$Z(\alpha)$  & $Z(\beta)$  & $Z(UR)$ \\
			\midrule
			\multirow{3}[2]{*}{\begin{sideways}{\tiny ICD-9-CM Vol 3.}\end{sideways}} & \multirow{3}[2]{*}{\begin{sideways}$<>$\quad\quad\end{sideways}} & \multirow{3}[2]{*}{\begin{sideways}{\tiny ICD-10-PCS}\quad\quad\end{sideways}} & $Z(\alpha)$  &       & \cellcolor[rgb]{ .627,  .816,  .498}0.99 & \cellcolor[rgb]{ .627,  .816,  .498}0.99 &       & \cellcolor[rgb]{ 1,  .922,  .518}0.97 & \cellcolor[rgb]{ 1,  .922,  .518}0.97 \\
			&       &       & $Z(\beta)$  & \cellcolor[rgb]{ .627,  .816,  .498}0.99 &       & \cellcolor[rgb]{ .388,  .745,  .482}1.00 & \cellcolor[rgb]{ 1,  .922,  .518}0.97 &       & \cellcolor[rgb]{ .392,  .749,  .486}1.00 \\
			&       &       & $Z(UR)$    & \cellcolor[rgb]{ .627,  .816,  .498}0.99 & \cellcolor[rgb]{ .388,  .745,  .482}1.00 &       & \cellcolor[rgb]{ 1,  .922,  .518}0.97 & \cellcolor[rgb]{ .392,  .749,  .486}1.00 &  \\
			\midrule
			\multirow{3}[1]{*}{\begin{sideways}{\tiny ICD-9-CM}\quad\quad\end{sideways}} & \multirow{3}[1]{*}{\begin{sideways}$<>$\quad\quad\end{sideways}} & \multirow{3}[1]{*}{\begin{sideways}{\tiny ICD-10-CM}\quad\quad\end{sideways}} & $Z(\alpha)$ &       & \cellcolor[rgb]{ .973,  .412,  .42}0.57 & \cellcolor[rgb]{ .992,  .804,  .494}0.88 &       & \cellcolor[rgb]{ .988,  .769,  .486}0.85 & \cellcolor[rgb]{ .992,  .796,  .494}0.87 \\
			&       &       & $Z(\beta)$  & \cellcolor[rgb]{ .973,  .412,  .42}0.57 &       & \cellcolor[rgb]{ .976,  .522,  .439}0.66 & \cellcolor[rgb]{ .988,  .769,  .486}0.85 &       & \cellcolor[rgb]{ .808,  .867,  .51}0.98 \\
			&       &       & $Z(UR)$    & \cellcolor[rgb]{ .992,  .804,  .494}0.88 & \cellcolor[rgb]{ .976,  .522,  .439}0.66 &       & \cellcolor[rgb]{ .992,  .796,  .494}0.87 & \cellcolor[rgb]{ .808,  .867,  .51}0.98 &  \\
		\end{tabular}%
	}
	\label{tab_corr}%
\end{table}%

\clearpage
\section{DISCUSSION}
In 2015, ICD-10-PCS had significantly more number of mapped procedure codes (n = 71,924), as compared to ICD-9-CM Vol.3 (n = 3,672) (see Table \ref{tab_procedure}). Equally, Table \ref{tab_diagnosis} shows more diagnosis codes for ICD-10-CM vis-à-vis ICD-9-CM. This fact alone implies that more specific information was likely to be gained by migrating from ICD-9-CM to ICD-10-CM/PCS, assuming complete clinical documentation and accurate coding. Besides, the mean statistics in these tables reveal that all the entropic measures are higher in the forward mappings, as compared to the backward mappings. This revelation further certifies that, on average, more information was gained in ICD-10-CM/PCS as compared to ICD-10-CM. The zero 25\%, 50\%, and 75\% quartile measures suggest that at least 75\% of codes in the source system had a one-to-one mapping with the target system. The implication of a one-to-one relationship is that no information is gained since $log (1) = 0$. In other words, the codes in a one-to-one mapping may structurally look different, but if they represent the same clinical concept, then no information is gained. To a computer, a one-to-one mapping is a simple translation, but, of course, to a human coder, more complicated code structures may be more challenging to extract and translate.

The scale of the information gained (or lost) between ICD-9-CM and ICD-10-CM/PCS can be appreciated by clinical classes depicted in Figures \ref{fig_procedure} and \ref{fig_diagnosis}. For example, Sub-figure \ref{fig_9-PCS-bar} indicates that in the procedural forward mappings, the most information was gained in the class of the \textit{Operations on Musculoskeletal System (76-84)}. The related box plot in Sub-figure \ref{fig_9-PCS-box} suggests that all three entropic measures relatively agreed on the characterization of class 76-84, given the small interquartile range. For diagnoses, Sub-figure \ref{fig_CM9-10-bar} suggests that the class of the \textit{Injury and Poisoning (800-999)} carried more information in ICD-10-CM followed by the class of \textit{Pregnancy and Childbirth (630-679)}.  Remarkably, Sub-figure \ref{fig_cm10-9-bar} implies that an ICD-10-CM class related to Pregnancy and Childbirth (O00-O9A) also resulted in information gain in ICD-9-CM. These conflicting results are due to the convoluted nature of the mappings between these two medical coding systems \cite{boyd2013discriminatory}.

From Figures \ref{fig_procedure} and \ref{fig_diagnosis}, it can be appreciated that some clinical classes have negative entropic measures. The implication of this observation is that little, or no information, was going to be gained in the target system. For example, regarding the procedures, Sub-figure \ref{fig_9-PCS-bar} indicates that for the class of the \textit{Diagnostic \& Therapeutic Procedures (87-99)}, little, or no information, was gained in ICD-10-PCS. Likewise, little, or no information, was going to be gained in ICD-9-CM Vol. 3 about the ICD-10-PCS class of \textit{Medical and Surgical (0)} (see  Sub-figure \ref{fig_PCS-9-bar}). Though, the entropic measures relatively disagree on the latter suggestion, given a large inter-quartile range of class (0) in Sub-figure \ref{fig_PCS-9-box}. Regarding diagnoses, Sub-figure \ref{fig_CM9-10-bar} suggests that little, or no information, was gained in ICD-10-CM about the ICD-9-CM class of \textit{Supplementary Classification Of External Causes Of Injury And Poisoning (E000-E999)}. In an apparent contraction, Sub-figure  \ref{fig_cm10-9-bar} points to little, or no information, gained in the backward mapping about ICD-10-CM classes of \textit{Injury, poisoning and certain other consequences of external causes (S00-T88)} and \textit{External causes of morbidity (V00-Y99)}. This ambiguity is again the result of convoluted mapping between ICD-9-CM and ICD-10-CM/PCS coding systems \cite{boyd2013discriminatory}.

It is noteworthy that, despite a greater number of codes in ICD-9-CM/PCS, the backward max statistics in both Tables \ref{tab_procedure} and \ref{tab_diagnosis} are not zero. The suggestion here is that, for some clinical concepts, ICD-9-CM captured more information vis-\`{a}-vis ICD-10-CM/PCS (e.g., class (F) in Sub-figure \ref{fig_PCS-9-bar} and class (O00-O9A) in Sub-figure \ref{fig_cm10-9-bar}).
The consequence of gaining information in the backward mapping is that some ICD-9-CM  information was lost in ICD-10-CM/PCS, which created issues with longitudinal data comparisons. This dilemma also likely produced problems with verifying ICD-10-CM/PCS codes' validity, especially for classes where the information was gained in both forward and backward mappings (bidirectional), such as in the pregnancy and childbirth clinical class. Additional challenges resulting from the bidirectional information gain include conflicting documentation requirements, especially if the new coding system is trying to collect different types of information than what is commonly documented. Naturally, coding errors are likely to result if clinical documentation is lacking.

To prepare for the transition to a new medical coding system, the user can utilize the proposed entropic measures as a guide to orient training efforts. To this end, clinical classes can be ranked as a way to gauge where most information is likely to be gained or lost. Of course, the user would have more confidence if the rankings of these entropic measures agreed. Regarding the transition from ICD-9-CM to ICD-10-CM/PCS, Table \ref{tab_corr} indicates that the most agreement existed in the forward mappings from ICD-10-CM Vol.3 to ICD-10-PCS. Another approach to prioritizing transition efforts may be based on the outlier and pattern analysis, as demonstrated in Figure \ref{fig_controls}. That is, instead of working with predefined clinical classes, the user could try to assess the impact of the transition using major themes from the descriptions of outlier maps. Many approaches to thematic analysis are available such as the Latent Dirichlet Allocation \cite{blei2003latent}. Here, a simple network was constructed and communities examined. For example, a close examination of Sub-figures \ref{fig_df31} and \ref{fig_df32} reveal a collection of terms that relate to various procedures for the vascular, skeletal, integumentary, and cardiac body systems. The most central words, in terms of the eigenvector centrality, were tissue, graft, subcutaneous, skin, repair, and incision. Combining these keywords, one may conclude that the procedures for the \textit{musculoskeletal, integumentary, and cardio-vascular systems} were likely to involve more information gain in ICD-10-PCS, a conclusion that is consistent with the results in Sub-figure \ref{fig_9-PCS-bar}.

\section{CONCLUSION} \label{sec_discussion}
Transitioning from an old medical coding system to a new one can be challenging, especially when the two coding systems are significantly different. The objective of this research was to propose methods that could help users prepare for the transition by identifying and focusing preparation initiatives on clinical concepts with more likelihood of transition challenges. To this end, two entropic measures of coding complexity were introduced. The first measure was a function of the variation in the alphabets of the codes in the target system, and the second measure was based on the possible number of valid representations of a code in the source system. It was recommended that the resulting entropic measures be normalized and adjusted by the probability of a given code before isolating clinical concepts of interest. The main assumption here is that the more entropy, the more likelihood of coding errors. So, more prudent documentation is required, not only to increase the chances of accurate coding but also code validity and longitudinal data comparisons. The proposed techniques are suitable for establishing transition complexity between any two medical coding systems, provided mappings or crosswalks exist. A demonstration of how to implement these techniques was carried out using the 2015 forward and backward mappings between  ICD-9-CM and ICD-10-CM/PCS.

\section{LIMITATIONS AND FUTURE RESEARCH} \label{sec_limitations}
A central conjecture of this research was that clinical concepts with more entropic measures were more likely to result in a more challenging transition. The justification of this assumption emanated from the fact that more entropic measures meant more variation in the codes, thus necessitating more prudent documentation and coding. This assumption may be violated if documentation is already complete and an experienced coder knows shortcuts to circumvent the new coding complexity. Accordingly, a medical record review may be necessary to ensure that the apparent complexity from the entropic measures actually exists. Besides, the topic of correlation between code validity and related entropic measures was not explored in this research. A medical record review may also be necessary to see if any lack of validity in the codes is explained by other reasons other than the entropic measures. Other relevant topics could not be considered in this research without additional data. For example, the question of how a new medical coding system could affect reimbursement was not entrained. Also, the topic of how coding guidelines and conventions may contribute to coding errors in the new system was not discussed. Future research goals include the consideration of these other topics, especially as they relate to the upcoming (or ongoing for some countries) transition from ICD-10 to ICD-11.

\clearpage
\appendix
\section{Calculating entropic measures of a map} \label{appendix_entropy}
\subsection{Calculating Shannon's entropy}
Given a source S with $m$ events and probabilities $p_1, p_2,\dots, p_m$ such that $\sum_{i= 1}^{m} p_i = 1$, the entropy $H$ of the source $S$ is computed as $H(S) = -\sum_{i = 1}^{m} p_i\log_2 p_i$, where $0\log 0 = 0$. The subscript of $2$ under the $\log$ symbol signifies bits units. The entropy of two independent sources $S$ and $T$ is given by $H(S, T) = H(S) + H(T)$. If the sources are dependent, the conditional entropy is used to obtain $H(S, T)$, as follows $H(S, T) = H(S) + H(T|S) = H(T) + H(S|T)$ \citep{luenberger}.

\subsection{Using Shannon's entropy to estimate coding complexity}
As noted in the main text, two major sources of coding complexity considered here are source $A$, which relates to the events of the alphabets in each column $\boldsymbol{a_{j}}$ of the map, for $j:1,\dots,n$, and source $B$, which relates to the events of the combinations of the rows (or codes) of the same map. Shannon's entropy of column $\boldsymbol{a_{j}}$, $H(\boldsymbol{a_{j}})$, is given by $H(\boldsymbol{a_{j}}) = -\sum_{i = 1}^{k_j} p_{ij}\log_2 p_{ij}$ where $k_j\leq m$ is the number of unique alphabets in column $\boldsymbol{a_{j}}$ and $p_{ij}$ is the probability of alphabet $i$ in position $j$. From Equation \ref{eq_H(A)}, it was indicated that $H(A) = \sum_{j=1}^{n} H(\boldsymbol{a_j}) = -\sum_{j = 1}^{n}\sum_{i = 1}^{k_j} p_{ij}\log_2 p_{ij}$. A more refined measure of $H(A)$ is possible to reflect the fact that the position of an alphabet in a code may carry a different weight of a coding error. For example, in ICD-10, a coding error where the first character is incorrect is typically much worse than a coding error where the last character is wrong since the first alphabet tends to serve as a root node for the classification of clinical concepts. Accordingly, a weighing scheme can be devised to account for the relative influence of the position of an alphabet in a code. If $w_j$ were the weight of position $j$ in the code, the resulting weighted average entropy of source A, $H(\bar{A})$, may look like this:
\begin{eqnarray} \label{eq_H(bar(A))}
H(\bar{A})= \frac{\sum_{j = 1}^{n} \log_2\left(\prod_{i=1}^{k_j}  p_{ij}^{- p_{ij}}\right)^{w_j}}{ \sum_{j = 1}^{n} w_j}
\end{eqnarray}
where, as before, $k_j\leq m$ is the number of unique alphabets in column $\boldsymbol{a_{j}}$ and $p_{ij}$ is the probability of alphabet $i$ in position $j$, for $j:1,\dots,n$. The proof of Equation \ref{eq_H(bar(A))} follows from a logarithmic rules of $\log(x.y) = \log(x) + \log(y) $ and $a.\log(x) = log(x)^a$. Hence, it follows that
$-\sum_{j = 1}^{n}\sum_{i = 1}^{k_j} p_{ij}\log p_{ij} = -\sum_{j = 1}^{n}\sum_{i = 1}^{k_j}  \log p_{ij}^{- p_{ij}} = \sum_{j = 1}^{n} \log\left( \prod_{i=i}^{k_j}  p_{ij}^{- p_{ij}}\right)$. After weighing each column, it is evident that $\sum_{j = 1}^{n} w_j \cdot\log\left( \prod_{i=i}^{k_j}  p_{ij}^{- p_{ij}}\right)$ is equivalent to  $\sum_{j = 1}^{n} \log\left(\prod_{i=1}^{k_j}  p_{ij}^{- p_{ij}}\right)^{w_j}$.  The denominator in Equation \ref{eq_H(bar(A))} allows for the calculation of the average of the weighted entropy.

As for source B, it was indicated in Equation \ref{eq_H(B)} that $H(B) =\log (v)$. The proof of this equation follows from the fact that, under the assumption of a uniform distribution, each valid representation is equally likely, with probability  $1/v$. Then by Shannon's entropy, $ H(B) = -\sum_{j= 1}^{v} \frac{1}{v} \log\left(\frac{1}{v}\right) = -\frac{v}{v} (\log 1 -\log(v)) = log(v)$. If all $m$ number of candidate codes in a map have a one-to-one relationship with code $x$, then all these candidate codes are considered stand-alone and don't have to be combined to form a valid representation of code $x$. In such case, the number of valid representations $v$ equals the number of stand-alone $m_0$, which also equals the number of candidate codes $m$. This means that:
\begin{eqnarray} \label{eq_m_0}
H(B) =  \log (v) \equiv  \log (m) \equiv  \log (m_0)
\end{eqnarray}
Equation \ref{eq_m_0} is comparable to a case of the UR measure in Chen et al. \cite{chen2018leveraging}. If a map includes some stand-alone codes and other codes that need to be combined under different scenarios to make valid representations of code $x$, then $v$ is obtained by:	
\begin{eqnarray} \label{eq_v}
v = m_0 + \sum_{i=1}^{s}[(m_1)(m_2)\dots (m_{m - m_0})]_i \equiv  m_0 + \sum_{i=1}^{s}\prod_{j = 1}^{m - m_0}m_{ij}
\end{eqnarray}
where $i$ represents scenario $i$ out of $s$ total number of scenarios. Here, $m_0$ is again the number of stand-alone codes and ${m - m_0}$ is the number of codes that must be combined in sequential order of their index, as a set, to represent the old code $x$. That is, for a given scenario $i$, $m_{i1}$ is the number of codes that must be sequenced first, followed by $m_{i2}$, the total number of codes that must be sequenced second followed by $m_{i3}$, the total number of codes that must be sequenced third, and so on until  $m_{i(m-m_0)}$.  If $m = m_0$, then, as indicated in Equation \ref{eq_m_0}, $v = m_0$. The justification of Equation \ref{eq_v} comes from the fact that if a map has some stand-alone codes, then there are ${m_0\choose 1} = \frac{m_0!}{1!(m_0-1)!} = m_0$ possibilities of choosing one stand-alone code at random. If a map has at least one scenario, then for each scenario $i$, there are ${m_{i1}\choose 1}.{m_{i2}\choose 1}\dots{m_{i(m - m_0)}\choose 1} = (m_{i1})(m_{i2})\dots(m_{i(m - m_0)})$ possibilities of choosing one sequence of codes from $m_{i1}$ to $m_{i(m - m_0)}$. Since stand-alone codes don't have to be combined with any other codes, then, for a map with a single scenario, the number of valid representations of code $x$ is given by $v = m_0 + (m_1)(m_2)\dots(m_{m - m_0})$ (review the example in Figure \ref{fig_example}). If a map had more than one scenario,  the total number of valid representations of code $x$ would follow Equation \ref{eq_v}.

\subsection{Normalizing the entropic measures}
Assuming that $H(A) \equiv \alpha$ and $H(B)\equiv \beta$, the normalized entropic scores can be obtained this way: 
\begin{eqnarray}\label{eq_normalize}
Z(\alpha) &=& \frac{\alpha - \bar{\alpha}}{var(\alpha)}\\\label{eq_normalize2}
Z(\beta) &=& \frac{\beta - \bar{\beta}}{var(\beta)}
\end{eqnarray} 
where $\bar{\alpha}$ is the average of $\alpha$ measures from all maps and $\bar{\beta}$ is the average of $\beta$ measures from all maps. The symbol $var()$ signifies the variance function.

\newpage
\section{Top 30 maps in the forward mapping from ICD-9-CM to ICD-10-CM}
\begin{table}[H]
	\centering
	\caption{Top 30 maps in the forward mapping from ICD-9-CM to ICD-10-CM, ranked by their sum of $Z(\alpha)$, $Z(\beta)$, and $Z(UR)$, from most to least total score. The map id and description correspond to ICD-9-CM code and description, respectively.}
	\scalebox{0.80}{
		\begin{tabular}{c|p{12cm}cccc}
			\textbf{Map} & \textbf{Map description} & \textbf{$Z(\alpha)$} & \textbf{$Z(\beta)$} & \textbf{$Z(UR)$} & \textbf{Total score} \\
		\midrule
		\textbf{V5412} & Aftercare for healing traumatic fracture of lower arm & 7.99  & 12.72 & 12.15 & \cellcolor[rgb]{ .388,  .745,  .482}32.87 \\
		\textbf{V5416} & Aftercare for healing traumatic fracture of lower leg & 7.95  & 12.47 & 11.90 & \cellcolor[rgb]{ .408,  .757,  .502}32.31 \\
		\textbf{V5411} & Aftercare for healing traumatic fracture of upper arm & 6.74  & 10.80 & 10.31 & \cellcolor[rgb]{ .569,  .82,  .635}27.86 \\
		\textbf{99529} & Unspecified adverse effect of other drug, medicinal and biological substance & 7.44  & 10.19 & 9.72  & \cellcolor[rgb]{ .584,  .827,  .651}27.36 \\
		\textbf{V5413} & Aftercare for healing traumatic fracture of hip & 7.14  & 10.19 & 9.72  & \cellcolor[rgb]{ .596,  .831,  .663}27.05 \\
		\textbf{V5417} & Aftercare for healing traumatic fracture of vertebrae & 6.78  & 10.00 & 9.54  & \cellcolor[rgb]{ .62,  .839,  .682}26.32 \\
		\textbf{V5415} & Aftercare for healing traumatic fracture of upper leg & 7.09  & 9.80  & 9.35  & \cellcolor[rgb]{ .624,  .843,  .686}26.23 \\
		\textbf{9895} & Toxic effect of venom & 6.12  & 10.10 & 9.63  & \cellcolor[rgb]{ .639,  .847,  .698}25.85 \\
		\textbf{99811} & Hemorrhage complicating a procedure & 10.01 & 8.08  & 7.70  & \cellcolor[rgb]{ .639,  .847,  .702}25.78 \\
		\textbf{99812} & Hematoma complicating a procedure & 10.01 & 8.08  & 7.70  & \cellcolor[rgb]{ .639,  .847,  .702}25.78 \\
		\textbf{V5419} & Aftercare for healing traumatic fracture of other bone & 7.73  & 9.11  & 8.69  & \cellcolor[rgb]{ .651,  .851,  .71}25.53 \\
		\textbf{29289} & Other specified drug-induced mental disorders & 6.43  & 9.34  & 8.91  & \cellcolor[rgb]{ .678,  .863,  .733}24.68 \\
		\textbf{9982} & Accidental puncture or laceration during a procedure, not elsewhere classified & 9.94  & 6.14  & 5.84  & \cellcolor[rgb]{ .776,  .902,  .82}21.92 \\
		\textbf{73382} & Nonunion of fracture & 6.45  & 6.90  & 6.56  & \cellcolor[rgb]{ .847,  .933,  .878}19.91 \\
		\textbf{9050} & Late effect of fracture of skull and face bones & 5.45  & 6.83  & 6.50  & \cellcolor[rgb]{ .886,  .949,  .914}18.78 \\
		\textbf{9947} & Asphyxiation and strangulation & 4.53  & 7.25  & 6.90  & \cellcolor[rgb]{ .89,  .949,  .918}18.68 \\
		\textbf{98989} & Toxic effect of other substance, chiefly nonmedicinal as to source, not elsewhere classified & 5.14  & 6.47  & 6.16  & \cellcolor[rgb]{ .922,  .965,  .945}17.77 \\
		\textbf{24980} & Secondary diabetes mellitus with other specified manifestations, not stated as uncontrolled, or unspecified & 4.93  & 6.47  & 6.16  & \cellcolor[rgb]{ .929,  .965,  .949}17.57 \\
		\textbf{9880} & Toxic effect of fish and shellfish eaten as food & 5.19  & 6.23  & 5.92  & \cellcolor[rgb]{ .937,  .969,  .957}17.34 \\
		\textbf{9823} & Toxic effect of other chlorinated hydrocarbon solvents & 4.36  & 6.55  & 6.23  & \cellcolor[rgb]{ .945,  .973,  .965}17.14 \\
		\textbf{9063} & Late effect of contusion & 6.13  & 5.63  & 5.35  & \cellcolor[rgb]{ .945,  .973,  .965}17.10 \\
		\textbf{8065} & Open fracture of lumbar spine with spinal cord injury & 5.13  & 5.63  & 5.92  & \cellcolor[rgb]{ .961,  .98,  .976}16.68 \\
		\textbf{9057} & Late effect of sprain and strain without mention of tendon injury & 6.16  & 5.38  & 5.11  & \cellcolor[rgb]{ .965,  .98,  .98}16.64 \\
		\textbf{E959} & Late effects of self-inflicted injury & 4.82  & 5.85  & 5.56  & \cellcolor[rgb]{ .976,  .984,  .992}16.22 \\
		\textbf{E9990} & Late effect of injury due to war operations & 4.81  & 5.85  & 5.56  & \cellcolor[rgb]{ .976,  .984,  .992}16.21 \\
		\textbf{64131} & Antepartum hemorrhage associated with coagulation defects, delivered, with or without mention of antepartum condition & 3.72  & 6.31  & 6.01  & \cellcolor[rgb]{ .984,  .988,  .996}16.04 \\
		\textbf{8064} & Closed fracture of lumbar spine with spinal cord injury & 4.49  & 5.63  & 5.92  & \cellcolor[rgb]{ .984,  .988,  .996}16.03 \\
		\textbf{9064} & Late effect of crushing & 5.50  & 5.38  & 5.11  & \cellcolor[rgb]{ .984,  .988,  1}15.98 \\
		\textbf{986} & Toxic effect of carbon monoxide & 4.56  & 5.85  & 5.56  & \cellcolor[rgb]{ .988,  .988,  1}15.97 \\
		\textbf{73395} & Stress fracture of other bone & 5.38  & 5.38  & 5.11  & \cellcolor[rgb]{ .988,  .988,  1}15.87 \\
		\end{tabular}%
	}
	\label{tab_icm9-icm10}%
\end{table}%

\section{Outliers maps of $Z(\alpha)$ where  $Z(\alpha) > 2.7$ (about top 1\% of the maps) }\label{appendix_za}

\begin{table}[H]
	\centering
	\caption{Outlier maps arranged in the decreasing order of the $Z(\alpha)$ scores. As before, $m$ is the number of candidate codes in a map  whereas $v$ is the number of valid representations in a map. The map id and description correspond to ICD-9-CM Vol. 3 code and description, respectively.}
	\scalebox{0.75}{
	\begin{tabular}{c|lccc}
		\textbf{Map} & \textbf{Map description} & \textbf{$m$} & \textbf{$v$} & \textbf{$\boldsymbol{Z(\alpha)}$} \\
		\midrule
		\textbf{3929} & other (peripheral) vascular shunt or bypass & 1191  & 1191  & \cellcolor[rgb]{ .388,  .745,  .482}4.87 \\
		\textbf{8605} & incision with removal of foreign body or device from skin and subcutaneous tissue & 415   & 415   & \cellcolor[rgb]{ .616,  .839,  .678}4.06 \\
		\textbf{8609} & other incision of skin and subcutaneous tissue & 335   & 335   & \cellcolor[rgb]{ .631,  .843,  .694}4.00 \\
		\textbf{3950} & angioplasty of other non-coronary vessel(s) & 1196  & 1196  & \cellcolor[rgb]{ .722,  .882,  .769}3.68 \\
		\textbf{0109} & other cranial puncture & 34    & 34    & \cellcolor[rgb]{ .722,  .882,  .769}3.68 \\
		\textbf{843} & revision of amputation stump & 349   & 349   & \cellcolor[rgb]{ .745,  .89,  .788}3.60 \\
		\textbf{9301} & functional evaluation & 148   & 148   & \cellcolor[rgb]{ .753,  .894,  .796}3.57 \\
		\textbf{0404} & other incision of cranial and peripheral nerves & 327   & 327   & \cellcolor[rgb]{ .78,  .906,  .82}3.47 \\
		\textbf{9788} & removal of external immobilization device & 98    & 98    & \cellcolor[rgb]{ .792,  .91,  .831}3.42 \\
		\textbf{3979} & other endovascular procedures on other vessels & 689   & 689   & \cellcolor[rgb]{ .796,  .914,  .835}3.41 \\
		\textbf{9223} & radioisotopic teleradiotherapy & 768   & 768   & \cellcolor[rgb]{ .804,  .914,  .843}3.38 \\
		\textbf{046} & transposition of cranial and peripheral nerves & 378   & 378   & \cellcolor[rgb]{ .808,  .918,  .843}3.37 \\
		\textbf{8382} & graft of muscle or fascia & 440   & 440   & \cellcolor[rgb]{ .812,  .918,  .847}3.35 \\
		\textbf{0124} & other craniotomy & 111   & 111   & \cellcolor[rgb]{ .827,  .925,  .863}3.29 \\
		\textbf{3926} & other intra-abdominal vascular shunt or bypass & 720   & 720   & \cellcolor[rgb]{ .827,  .925,  .863}3.29 \\
		\textbf{409} & other operations on lymphatic structures & 510   & 510   & \cellcolor[rgb]{ .835,  .929,  .871}3.27 \\
		\textbf{8196} & other repair of joint & 313   & 313   & \cellcolor[rgb]{ .839,  .929,  .871}3.26 \\
		\textbf{9227} & implantation or insertion of radioactive elements & 268   & 268   & \cellcolor[rgb]{ .855,  .933,  .882}3.21 \\
		\textbf{3897} & central venous catheter placement with guidance & 48    & 320   & \cellcolor[rgb]{ .871,  .941,  .898}3.15 \\
		\textbf{0474} & other anastomosis of cranial or peripheral nerve & 350   & 350   & \cellcolor[rgb]{ .871,  .941,  .898}3.15 \\
		\textbf{3821} & biopsy of blood vessel & 699   & 699   & \cellcolor[rgb]{ .89,  .949,  .914}3.08 \\
		\textbf{8120} & arthrodesis of unspecified joint & 582   & 582   & \cellcolor[rgb]{ .894,  .953,  .918}3.06 \\
		\textbf{3958} & repair of blood vessel with unspecified type of patch graft & 421   & 421   & \cellcolor[rgb]{ .894,  .953,  .922}3.05 \\
		\textbf{8604} & other incision with drainage of skin and subcutaneous tissue & 281   & 281   & \cellcolor[rgb]{ .906,  .957,  .929}3.01 \\
		\textbf{3348} & other repair and plastic operations on bronchus & 288   & 288   & \cellcolor[rgb]{ .91,  .957,  .933}3.00 \\
		\textbf{8129} & arthrodesis of other specified joints & 552   & 552   & \cellcolor[rgb]{ .91,  .957,  .933}3.00 \\
		\textbf{2103} & control of epistaxis by cauterization (and packing) & 10    & 8     & \cellcolor[rgb]{ .91,  .957,  .933}3.00 \\
		\textbf{9649} & other genitourinary instillation & 5     & 5     & \cellcolor[rgb]{ .918,  .961,  .941}2.97 \\
		\textbf{3956} & repair of blood vessel with tissue patch graft & 402   & 402   & \cellcolor[rgb]{ .925,  .965,  .945}2.95 \\
		\textbf{8100} & spinal fusion, not otherwise specified & 282   & 282   & \cellcolor[rgb]{ .945,  .973,  .965}2.88 \\
		\textbf{8683} & size reduction plastic operation & 340   & 340   & \cellcolor[rgb]{ .957,  .976,  .973}2.84 \\
		\textbf{9205} & cardiovascular and hematopoietic scan and radioisotope function study & 54    & 54    & \cellcolor[rgb]{ .961,  .976,  .976}2.83 \\
		\textbf{0309} & other exploration and decompression of spinal canal & 70    & 70    & \cellcolor[rgb]{ .973,  .984,  .988}2.78 \\
		\textbf{8080} & other local excision or destruction of lesion of joint, unspecified site & 345   & 345   & \cellcolor[rgb]{ .976,  .984,  .988}2.77 \\
		\textbf{8040} & division of joint capsule, ligament, or cartilage, unspecified site & 210   & 210   & \cellcolor[rgb]{ .984,  .988,  .996}2.74 \\
		\textbf{3925} & aorta-iliac-femoral bypass & 320   & 320   & \cellcolor[rgb]{ .984,  .988,  .996}2.73 \\
		\textbf{3805} & incision of vessel, other thoracic vessels & 78    & 78    & \cellcolor[rgb]{ .988,  .988,  1}2.71 \\
	\end{tabular}%
}
\label{tab_outlirs_za}%
\end{table}%

\section{Outliers maps of $Z(\beta)$ where  $Z(\beta) > 2.85$ (about top 1\% of the maps)}\label{appendix_zb}

\begin{table}[H]
	\centering
	\caption{Outlier maps arranged in the decreasing order of the $Z(\beta)$ scores. As before, $m$ is the number of candidate codes in a map  whereas $v$ is the number of valid representations in a map. The $UR$ measure (obtained by $\log_2(m)$) can directly  be compared to $H(B)$ (obtained by $\log_2(v)$). When $UR < H(B)$, $UR$ has underestimated the expected coding complexity. When $UR > H(B)$, $UR$ has overestimated the expected coding complexity. Otherwise, the measures are equal. The map id and description correspond to ICD-9-CM code and description, respectively.}
	\scalebox{0.75}{
	\begin{tabular}{c|lccccc}
		\textbf{Map} & \textbf{Map description} & \textbf{$H(B)$} & \textbf{$UR$} & \textbf{$m$} & \textbf{$v$} & \textbf{$\boldsymbol{Z(\beta)}$} \\
		\midrule
		\textbf{3473} & Closure of other fistula of thorax & 10.95 & 7.92  & 243   & 1977  & \cellcolor[rgb]{ .388,  .745,  .482}4.26 \\
		\textbf{3950} & Angioplasty of other non-coronary vessel(s) & 10.22 & 10.22 & 1196  & 1196  & \cellcolor[rgb]{ .553,  .812,  .624}3.88 \\
		\textbf{3929} & Other (peripheral) vascular shunt or bypass & 10.22 & 10.22 & 1191  & 1191  & \cellcolor[rgb]{ .553,  .816,  .627}3.88 \\
		\textbf{304} & Radical laryngectomy & 9.61  & 5.81  & 56    & 784   & \cellcolor[rgb]{ .69,  .871,  .745}3.57 \\
		\textbf{9223} & Radioisotopic teleradiotherapy & 9.58  & 9.58  & 768   & 768   & \cellcolor[rgb]{ .698,  .871,  .749}3.55 \\
		\textbf{3926} & Other intra-abdominal vascular shunt or bypass & 9.49  & 9.49  & 720   & 720   & \cellcolor[rgb]{ .718,  .878,  .769}3.50 \\
		\textbf{3821} & Biopsy of blood vessel & 9.45  & 9.45  & 699   & 699   & \cellcolor[rgb]{ .729,  .882,  .776}3.48 \\
		\textbf{3979} & Other endovascular procedures on other vessels & 9.43  & 9.43  & 689   & 689   & \cellcolor[rgb]{ .733,  .886,  .78}3.47 \\
		\textbf{8120} & Arthrodesis of unspecified joint & 9.18  & 9.18  & 582   & 582   & \cellcolor[rgb]{ .788,  .91,  .827}3.34 \\
		\textbf{5684} & Closure of other fistula of ureter & 9.13  & 6.13  & 70    & 560   & \cellcolor[rgb]{ .8,  .914,  .839}3.31 \\
		\textbf{8129} & Arthrodesis of other specified joints & 9.11  & 9.11  & 552   & 552   & \cellcolor[rgb]{ .804,  .914,  .843}3.30 \\
		\textbf{409} & Other operations on lymphatic structures & 8.99  & 8.99  & 510   & 510   & \cellcolor[rgb]{ .831,  .925,  .863}3.24 \\
		\textbf{8687} & Fat graft of skin and subcutaneous tissue & 8.98  & 6.04  & 66    & 506   & \cellcolor[rgb]{ .831,  .925,  .867}3.24 \\
		\textbf{5783} & Repair of fistula involving bladder and intestine & 8.98  & 6.13  & 70    & 505   & \cellcolor[rgb]{ .835,  .925,  .867}3.23 \\
		\textbf{3342} & Closure of bronchial fistula & 8.78  & 7.03  & 131   & 440   & \cellcolor[rgb]{ .878,  .945,  .906}3.13 \\
		\textbf{5784} & Repair of other fistula of bladder & 8.78  & 6.19  & 73    & 440   & \cellcolor[rgb]{ .878,  .945,  .906}3.13 \\
		\textbf{8382} & Graft of muscle or fascia & 8.78  & 8.78  & 440   & 440   & \cellcolor[rgb]{ .878,  .945,  .906}3.13 \\
		\textbf{3794} & Implantation or replacement of automatic cardioverter/defibrillator, total system [aicd] & 8.75  & 5.25  & 38    & 432   & \cellcolor[rgb]{ .886,  .949,  .91}3.12 \\
		\textbf{3958} & Repair of blood vessel with unspecified type of patch graft & 8.72  & 8.72  & 421   & 421   & \cellcolor[rgb]{ .894,  .949,  .918}3.10 \\
		\textbf{8605} & Incision with removal of foreign body or device from skin and subcutaneous tissue & 8.70  & 8.70  & 415   & 415   & \cellcolor[rgb]{ .898,  .953,  .922}3.09 \\
		\textbf{3956} & Repair of blood vessel with tissue patch graft & 8.65  & 8.65  & 402   & 402   & \cellcolor[rgb]{ .906,  .957,  .929}3.06 \\
		\textbf{046} & Transposition of cranial and peripheral nerves & 8.56  & 8.56  & 378   & 378   & \cellcolor[rgb]{ .929,  .965,  .949}3.02 \\
		\textbf{8663} & Full-thickness skin graft to other sites & 8.56  & 5.81  & 56    & 378   & \cellcolor[rgb]{ .929,  .965,  .949}3.02 \\
		\textbf{3991} & Freeing of vessel & 8.49  & 8.49  & 360   & 360   & \cellcolor[rgb]{ .945,  .973,  .961}2.98 \\
		\textbf{3957} & Repair of blood vessel with synthetic patch graft & 8.48  & 8.48  & 358   & 358   & \cellcolor[rgb]{ .945,  .973,  .965}2.98 \\
		\textbf{3951} & Clipping of aneurysm & 8.48  & 8.48  & 357   & 357   & \cellcolor[rgb]{ .945,  .973,  .965}2.97 \\
		\textbf{0474} & Other anastomosis of cranial or peripheral nerve & 8.45  & 8.45  & 350   & 350   & \cellcolor[rgb]{ .953,  .976,  .969}2.96 \\
		\textbf{843} & Revision of amputation stump & 8.45  & 8.45  & 349   & 349   & \cellcolor[rgb]{ .953,  .976,  .969}2.96 \\
		\textbf{3959} & Other repair of vessel & 8.43  & 8.43  & 346   & 346   & \cellcolor[rgb]{ .957,  .976,  .973}2.95 \\
		\textbf{8080} & Other local excision or destruction of lesion of joint, unspecified site & 8.43  & 8.43  & 345   & 345   & \cellcolor[rgb]{ .957,  .976,  .973}2.95 \\
		\textbf{8683} & Size reduction plastic operation & 8.41  & 8.41  & 340   & 340   & \cellcolor[rgb]{ .961,  .98,  .976}2.94 \\
		\textbf{8609} & Other incision of skin and subcutaneous tissue & 8.39  & 8.39  & 335   & 335   & \cellcolor[rgb]{ .969,  .98,  .98}2.93 \\
		\textbf{0404} & Other incision of cranial and peripheral nerves & 8.35  & 8.35  & 327   & 327   & \cellcolor[rgb]{ .976,  .984,  .988}2.91 \\
		\textbf{3897} & Central venous catheter placement with guidance & 8.32  & 5.58  & 48    & 320   & \cellcolor[rgb]{ .98,  .988,  .996}2.89 \\
		\textbf{3925} & Aorta-iliac-femoral bypass & 8.32  & 8.32  & 320   & 320   & \cellcolor[rgb]{ .98,  .988,  .996}2.89 \\
		\textbf{8196} & Other repair of joint & 8.29  & 8.29  & 313   & 313   & \cellcolor[rgb]{ .988,  .988,  1}2.88 \\
		\textbf{8381} & Tendon graft & 8.29  & 8.29  & 312   & 312   & \cellcolor[rgb]{ .988,  .988,  1}2.87 \\
		\textbf{8383} & Tendon pulley reconstruction other than hand & 8.29  & 8.29  & 312   & 312   & \cellcolor[rgb]{ .988,  .988,  1}2.87 \\
		\end{tabular}%
		}
\label{tab_outlirs_zb}%
\end{table}%

%


\newpage
\section{Implementation of Algorithm \ref{algorithm_shannon} using Python 3.6}\label{appendix_python}
To download the gem\_i9pcs.txt file used in this code - this file represents the forward mappings from ICD-9-CM Vol.3 to ICD-10-PCS, go to \url{https://www.cms.gov/Medicare/Coding/ICD10/2015-ICD-10-PCS-and-GEMs}$>$\textit{2015 General Equivalence Mappings (GEMs) – Procedure Codes and Guide (ZIP)} $>$\textit{gem\_i9pcs.txt}
\begin{lstlisting}[language=Python]
import pandas as pd
from math import log
from collections import Counter
import numpy as np
##Import forward mappings, ICD-10-CM Vol 3 to ICD-10-PCS. If importing backward mappings, add a padding to icd9 codes so all candidate codes have a fixed length (e.g., use lambda x:x.ljust(7, '0') to have all codes be 7 characters long). Also, for backward mappings, ensure to replace all instances of pcs (in this script) with icd9.
icd10=pd.read_csv(r'path\gem_i9pcs.txt',sep='\s+',names=['icd9','pcs','flag'],converters={'flag': lambda x: str(x),'icd9': lambda x: str(x) })
icd10 = icd10[~icd10.pcs.isin(['NoPCS', 'NoDx'])]#Exclude cases with no match in the target system
##Relevant functions to implement Algorithm 4.1
def slice_code(code,n): 
    '''slice a code into axes(e.g., [ABC]-->[A,B,C]) where n is the constant code length in a map''' 
    if code =='NoPCS':
        pass
    else:
        return code[n]
def count_flag(code):    
    ''' Count the number of possible combinations, given the flag column in icd10 file'''    
    b = pd.Series(code).value_counts().sort_index()
    b = pd.Series(b)
    c = list(b.index)
    d = list(b.array)
    g=[]
    for i in b.index:    
        g.append (int(str(i)[3]))
    s1 = pd.Series(d)
    s2 = pd.Series(g)
    s3 = pd.concat([s2,s1], axis=1)
    s4 = s3.groupby(0)[1].apply(np.prod)
    s4 = pd.Series(s4)
    s5 = sum(list(s4.array))
    return s5
def entropy(aj): 
    '''calculate the entropy of each axis (aj) to obtain H(a_j)'''
    p, lns = Counter(aj), float(len(aj))
    return abs(round(-sum( ((count/lns)*log(count/lns, 2))  for count in p.values()),2)) 
def log_function(x):
    '''calculate the entropy, given v (number of possible combinations) to obtain H(B)'''
    if x==0:
        return 0
    else:
        return log (x,2)
##STEP 1: CALCULATE H(A)
#Slice the pcs code
icd10['section']=icd10.pcs.apply(slice_code, n=0)
icd10['system']=icd10.pcs.apply(slice_code, n=1)
icd10['root']=icd10.pcs.apply(slice_code, n=2)
icd10['part']=icd10.pcs.apply(slice_code, n=3)
icd10['approach']=icd10.pcs.apply(slice_code, n=4)
icd10['device']=icd10.pcs.apply(slice_code, n=5)
icd10['qualifier']=icd10.pcs.apply(slice_code, n=6)
#Calculate H(aj)
axis1 = icd10.groupby('icd9')['section'].apply(entropy)
axis2  = icd10.groupby('icd9')['system'].apply(entropy)
axis3  = icd10.groupby('icd9')['root'].apply(entropy)
axis4  = icd10.groupby('icd9')['part'].apply(entropy)
axis5  = icd10.groupby('icd9')['approach'].apply(entropy)
axis6  = icd10.groupby('icd9')['device'].apply(entropy)
axis7  = icd10.groupby('icd9')['qualifier'].apply(entropy)
#Aggreate a dataframe of all column entropic measures
maps = pd.concat([axis1,axis2,axis3,axis4,axis5,axis6,axis7], axis =1)
maps = pd.DataFrame(maps)
maps.columns = ['H(a1)','H(a2)','H(a3)','H(a4)','H(a5)','H(a6)','H(a7)']
#Equal weights for axes: np.array([1,1,1,1,1,1,1]). H(A)=sum(H(aj)) 
weights = np.array([1,1,1,1,1,1,1]) 
maps['H(A)'] = np.dot(maps, weights)
##STEP 2: CALCULATE H(B) AND UR
maps['v'] = icd10.groupby('icd9')['flag'].apply(count_flag)
maps['H(B)'] = maps['v'].apply(lambda x: log_function(x))
maps['m'] = icd10.groupby('icd9')['flag'].count()
maps['UR'] = maps['m'].apply(lambda x: log_function(x))
##STEP 3: NORMALIZE ENTROPIC MEASURES
#Normalize H(A)
avg = maps['H(A)'].mean()
var = maps['H(A)'].var()
maps['Z(alpha)'] = maps['H(A)'].apply(lambda x: (x-avg)/np.sqrt(var))
#Normalize H(B)
avg = maps['H(B)'].mean()
var = maps['H(B)'].var()
maps['Z(beta)'] = maps['H(B)'].apply(lambda x: (x-avg)/np.sqrt(var))
#Normalize the UR measure
avg = maps['UR'].mean()
var = maps['UR'].var()
maps['Z(UR)'] = maps['UR'].apply(lambda x: (x-avg)/np.sqrt(var))
##STEP 4: WEIGH ENTROPIC MEASURES by the historical frequency distribution of clincal concepts (if data available)
##STEP 5: RANK MAPS from highest to lowest entropic measure
maps['Zsum'] = maps['Z(alpha)']+maps['Z(beta)']+maps['Z(UR)']
##maps_rankall: rank maps by combining all entropic measures
maps_rankall = maps.sort_values(by=['Zsum'],  ascending=False)
##maps_rank1: rank maps by Z(alpha)
maps_rank1 = maps.sort_values(by=['Z(alpha)'],  ascending=False)
##maps_rank2: rank maps by Z(beta)
maps_rank2 = maps.sort_values(by=['Z(beta)'],  ascending=False)
##maps_rank3: rank maps y Z(UR)
maps_rank3 = maps.sort_values(by=['Z(UR)'],  ascending=False)

\end{lstlisting}	

%
%
%
%
%
%
\newpage
\bibliography{bibfile}

\begin{thebibliography}{24}
\expandafter\ifx\csname natexlab\endcsname\relax\def\natexlab#1{#1}\fi
\providecommand{\url}[1]{\texttt{#1}}
\providecommand{\href}[2]{#2}
\providecommand{\path}[1]{#1}
\providecommand{\DOIprefix}{doi:}
\providecommand{\ArXivprefix}{arXiv:}
\providecommand{\URLprefix}{URL: }
\providecommand{\Pubmedprefix}{pmid:}
\providecommand{\doi}[1]{\href{http://dx.doi.org/#1}{\path{#1}}}
\providecommand{\Pubmed}[1]{\href{pmid:#1}{\path{#1}}}
\providecommand{\bibinfo}[2]{#2}
\ifx\xfnm\relax \def\xfnm[#1]{\unskip,\space#1}\fi
\bibitem[{CMS(2014)}]{cms_guidelines}
\bibinfo{author}{CMS}, \bibinfo{title}{{2015 Official ICD-10-PCS Coding
  Guidelines}}, \bibinfo{year}{2014}. \URLprefix
  \url{http://www.cms.gov/Medicare/Coding/ICD10/Downloads/PCS-2014-guidelines.pdf}.
\bibitem[{NCHS(2014)}]{nchs}
\bibinfo{author}{NCHS}, \bibinfo{title}{{International Classification of
  Diseases, Tenth Revision, Clinical Modification (ICD-10-CM)}},
  \bibinfo{year}{2014}. \URLprefix
  \url{http://www.cdc.gov/nchs/icd/icd10cm.htm}.
\bibitem[{Caskey et~al.(2014)Caskey, Zaman, Nam, Chae, Williams, Mathew,
  Burton, Lussier, Boyd et~al.}]{caskey2014transition}
\bibinfo{author}{R.~Caskey}, \bibinfo{author}{J.~Zaman},
  \bibinfo{author}{H.~Nam}, \bibinfo{author}{S.-R. Chae},
  \bibinfo{author}{L.~Williams}, \bibinfo{author}{G.~Mathew},
  \bibinfo{author}{M.~Burton}, \bibinfo{author}{Y.~A. Lussier},
  \bibinfo{author}{A.~D. Boyd}, et~al.,
\newblock \bibinfo{title}{The transition to icd-10-cm: challenges for pediatric
  practice},
\newblock \bibinfo{journal}{Pediatrics} \bibinfo{volume}{134}
  (\bibinfo{year}{2014}) \bibinfo{pages}{31--36}.
\bibitem[{Kusnoor et~al.(2019)Kusnoor, Blasingame, Williams, DesAutels, Su, and
  Giuse}]{kusnoor2019narrative}
\bibinfo{author}{S.~V. Kusnoor}, \bibinfo{author}{M.~N. Blasingame},
  \bibinfo{author}{A.~M. Williams}, \bibinfo{author}{S.~J. DesAutels},
  \bibinfo{author}{J.~Su}, \bibinfo{author}{N.~B. Giuse},
\newblock \bibinfo{title}{A narrative review of the impact of the transition to
  icd-10 and icd-10-cm/pcs},
\newblock \bibinfo{journal}{JAMIA Open}  (\bibinfo{year}{2019}).
\bibitem[{Alakrawi et~al.(2017)Alakrawi, Watzlaf, Nemchik, and
  Sheridan}]{alakrawi2017new}
\bibinfo{author}{Z.~M. Alakrawi}, \bibinfo{author}{V.~Watzlaf},
  \bibinfo{author}{S.~Nemchik}, \bibinfo{author}{P.~T. Sheridan},
\newblock \bibinfo{title}{New study illuminates the ongoing road to icd-10
  productivity and optimization},
\newblock \bibinfo{journal}{J AHIMA} \bibinfo{volume}{88}
  (\bibinfo{year}{2017}) \bibinfo{pages}{40--45}.
\bibitem[{Monestime et~al.(2019)Monestime, Mayer, and
  Blackwood}]{monestime2019analyzing}
\bibinfo{author}{J.~P. Monestime}, \bibinfo{author}{R.~W. Mayer},
  \bibinfo{author}{A.~Blackwood},
\newblock \bibinfo{title}{Analyzing the icd-10-cm transition and
  post-implementation stages: A public health institution case study},
\newblock \bibinfo{journal}{Perspectives in health information management}
  \bibinfo{volume}{16} (\bibinfo{year}{2019}).
\bibitem[{Butler(2016)}]{butler2016analyzing}
\bibinfo{author}{M.~Butler},
\newblock \bibinfo{title}{Analyzing eight months of icd-10},
\newblock \bibinfo{journal}{Journal of AHIMA} \bibinfo{volume}{87}
  (\bibinfo{year}{2016}) \bibinfo{pages}{15--22}.
\bibitem[{Stitcher and Lawrence(2016)}]{Stitcher}
\bibinfo{author}{S.~Stitcher}, \bibinfo{author}{H.~Lawrence},
  \bibinfo{title}{From icd-9 to icd10: A comparative analysis of coding audit
  findings in year one}, \bibinfo{year}{2016}. \URLprefix
  \url{https://hfmamd.starchapter.com/downloads/46TH_ANNUAL_INSTITUTE/3__horizon_presentation__hfma_2016_sep22.pptx}.
\bibitem[{Hellman et~al.(2018)Hellman, Lim, Leung, Blount, and
  Yiu}]{hellman2018impact}
\bibinfo{author}{J.~B. Hellman}, \bibinfo{author}{M.~C. Lim},
  \bibinfo{author}{K.~Y. Leung}, \bibinfo{author}{C.~M. Blount},
  \bibinfo{author}{G.~Yiu},
\newblock \bibinfo{title}{The impact of conversion to international
  classification of diseases, 10th revision (icd-10) on an academic
  ophthalmology practice},
\newblock \bibinfo{journal}{Clinical ophthalmology (Auckland, NZ)}
  \bibinfo{volume}{12} (\bibinfo{year}{2018}) \bibinfo{pages}{949}.
\bibitem[{Krive et~al.(2015)Krive, Patel, Gehm, Mackey, Kulstad, Lussier, Boyd
  et~al.}]{krive2015complexity}
\bibinfo{author}{J.~Krive}, \bibinfo{author}{M.~Patel},
  \bibinfo{author}{L.~Gehm}, \bibinfo{author}{M.~Mackey},
  \bibinfo{author}{E.~Kulstad}, \bibinfo{author}{Y.~A. Lussier},
  \bibinfo{author}{A.~D. Boyd}, et~al.,
\newblock \bibinfo{title}{The complexity and challenges of the international
  classification of diseases, ninth revision, clinical modification to
  international classification of diseases, 10th revision, clinical
  modification transition in eds},
\newblock \bibinfo{journal}{The American journal of emergency medicine}
  \bibinfo{volume}{33} (\bibinfo{year}{2015}) \bibinfo{pages}{713--718}.
\bibitem[{Januel et~al.(2011)Januel, Luthi, Quan, Borst, Taff{\'e}, Ghali, and
  Burnand}]{januel2011improved}
\bibinfo{author}{J.-M. Januel}, \bibinfo{author}{J.-C. Luthi},
  \bibinfo{author}{H.~Quan}, \bibinfo{author}{F.~Borst},
  \bibinfo{author}{P.~Taff{\'e}}, \bibinfo{author}{W.~A. Ghali},
  \bibinfo{author}{B.~Burnand},
\newblock \bibinfo{title}{Improved accuracy of co-morbidity coding over time
  after the introduction of icd-10 administrative data},
\newblock \bibinfo{journal}{BMC health services research} \bibinfo{volume}{11}
  (\bibinfo{year}{2011}) \bibinfo{pages}{194}.
\bibitem[{Quan et~al.(2008)Quan, Li, Duncan~Saunders, Parsons, Nilsson,
  Alibhai, Ghali, and investigators}]{quan2008assessing}
\bibinfo{author}{H.~Quan}, \bibinfo{author}{B.~Li},
  \bibinfo{author}{L.~Duncan~Saunders}, \bibinfo{author}{G.~A. Parsons},
  \bibinfo{author}{C.~I. Nilsson}, \bibinfo{author}{A.~Alibhai},
  \bibinfo{author}{W.~A. Ghali}, \bibinfo{author}{I.~investigators},
\newblock \bibinfo{title}{Assessing validity of icd-9-cm and icd-10
  administrative data in recording clinical conditions in a unique dually coded
  database},
\newblock \bibinfo{journal}{Health services research} \bibinfo{volume}{43}
  (\bibinfo{year}{2008}) \bibinfo{pages}{1424--1441}.
\bibitem[{WHO(2020)}]{who_icd11}
\bibinfo{author}{WHO}, \bibinfo{title}{Who releases new international
  classification of diseases (icd 11)}, \bibinfo{year}{2020}. \URLprefix
  \url{https://www.who.int/news-room/detail/18-06-2018-who-releases-new-international-classification-of-diseases-(icd-11)}.
\bibitem[{CDC(2018)}]{cdc_icd11}
\bibinfo{author}{CDC}, \bibinfo{title}{Update on icd-11: The who launch and
  implications for u.s. implementation}, \bibinfo{year}{2018}. \URLprefix
  \url{https://www.cdc.gov/nchs/data/icd/ICD-11-WHOV-CM-2018-V3.pdf}.
\bibitem[{Fung et~al.(2020)Fung, Xu, and Bodenreider}]{fung2020new}
\bibinfo{author}{K.~W. Fung}, \bibinfo{author}{J.~Xu},
  \bibinfo{author}{O.~Bodenreider},
\newblock \bibinfo{title}{The new international classification of diseases 11th
  edition: a comparative analysis with icd-10 and icd-10-cm},
\newblock \bibinfo{journal}{Journal of the American Medical Informatics
  Association} \bibinfo{volume}{27} (\bibinfo{year}{2020})
  \bibinfo{pages}{738--746}.
\bibitem[{Boyd et~al.(2013)Boyd, Li, Burton, Jonen, Gardeux, Achour, Luo,
  Zenku, Bahroos, Brown et~al.}]{boyd2013discriminatory}
\bibinfo{author}{A.~D. Boyd}, \bibinfo{author}{J.~J. Li},
  \bibinfo{author}{M.~D. Burton}, \bibinfo{author}{M.~Jonen},
  \bibinfo{author}{V.~Gardeux}, \bibinfo{author}{I.~Achour},
  \bibinfo{author}{R.~Q. Luo}, \bibinfo{author}{I.~Zenku},
  \bibinfo{author}{N.~Bahroos}, \bibinfo{author}{S.~B. Brown}, et~al.,
\newblock \bibinfo{title}{The discriminatory cost of icd-10-cm transition
  between clinical specialties: metrics, case study, and mitigating tools},
\newblock \bibinfo{journal}{Journal of the American Medical Informatics
  Association} \bibinfo{volume}{20} (\bibinfo{year}{2013})
  \bibinfo{pages}{708--717}.
\bibitem[{Boyd et~al.(2018)Boyd, Jianrong‘John’Li, Zaim, Krive, Mittal,
  Satava, Burton, Smith, and Lussier}]{boyd2018icd}
\bibinfo{author}{A.~D. Boyd}, \bibinfo{author}{C.~K. Jianrong‘John’Li},
  \bibinfo{author}{S.~R. Zaim}, \bibinfo{author}{J.~Krive},
  \bibinfo{author}{M.~Mittal}, \bibinfo{author}{R.~A. Satava},
  \bibinfo{author}{M.~Burton}, \bibinfo{author}{J.~Smith},
  \bibinfo{author}{Y.~A. Lussier},
\newblock \bibinfo{title}{Icd-10 procedure codes produce transition
  challenges},
\newblock \bibinfo{journal}{AMIA Summits on Translational Science Proceedings}
  \bibinfo{volume}{2018} (\bibinfo{year}{2018}) \bibinfo{pages}{35}.
\bibitem[{Chen et~al.(2018)Chen, Zhang, and Zhu}]{chen2018leveraging}
\bibinfo{author}{D.~Chen}, \bibinfo{author}{R.~Zhang},
  \bibinfo{author}{X.~Zhu},
\newblock \bibinfo{title}{Leveraging shannon entropy to validate the transition
  between icd-10 and icd-11},
\newblock \bibinfo{journal}{Entropy} \bibinfo{volume}{20}
  (\bibinfo{year}{2018}) \bibinfo{pages}{769}.
\bibitem[{Luenberger(2006)}]{luenberger}
\bibinfo{author}{D.~G. Luenberger}, \bibinfo{title}{Information Science},
  \bibinfo{publisher}{Princeton University Press}, \bibinfo{year}{2006}.
\bibitem[{CMS(2014)}]{cms_gem}
\bibinfo{author}{CMS}, \bibinfo{title}{{Procedure Code Set General Equivalence
  Mappings ICD-10-PCS to ICD-9-CM and ICD-9-CM to ICD-10-PCS Documentation and
  User’s Guide}}, \bibinfo{year}{2014}. \URLprefix
  \url{http://www.cms.gov/Medicare/Coding/ICD10/2015-ICD-10-CM-and-GEMs.html}.
\bibitem[{Niyirora and Aragones(2020)}]{niyirora2019network}
\bibinfo{author}{J.~Niyirora}, \bibinfo{author}{O.~Aragones},
\newblock \bibinfo{title}{Network analysis of medical care services},
\newblock \bibinfo{journal}{Health informatics journal} \bibinfo{volume}{26}
  (\bibinfo{year}{2020}) \bibinfo{pages}{1631--1658}.
\bibitem[{Wilbur and Sirotkin(1992)}]{wilbur1992automatic}
\bibinfo{author}{W.~J. Wilbur}, \bibinfo{author}{K.~Sirotkin},
\newblock \bibinfo{title}{The automatic identification of stop words},
\newblock \bibinfo{journal}{Journal of information science}
  \bibinfo{volume}{18} (\bibinfo{year}{1992}) \bibinfo{pages}{45--55}.
\bibitem[{Bastian et~al.(2009)Bastian, Heymann, and Jacomy}]{bastian2009gephi}
\bibinfo{author}{M.~Bastian}, \bibinfo{author}{S.~Heymann},
  \bibinfo{author}{M.~Jacomy},
\newblock \bibinfo{title}{Gephi: an open source software for exploring and
  manipulating networks},
\newblock in: \bibinfo{booktitle}{Third international AAAI conference on
  weblogs and social media}, \bibinfo{year}{2009}.
\bibitem[{Blei et~al.(2003)Blei, Ng, and Jordan}]{blei2003latent}
\bibinfo{author}{D.~M. Blei}, \bibinfo{author}{A.~Y. Ng},
  \bibinfo{author}{M.~I. Jordan},
\newblock \bibinfo{title}{Latent dirichlet allocation},
\newblock \bibinfo{journal}{Journal of machine Learning research}
  \bibinfo{volume}{3} (\bibinfo{year}{2003}) \bibinfo{pages}{993--1022}.

\end{thebibliography}

\end{document}